\documentclass[
 preprint,
 superscriptaddress,
 amsmath,amssymb,
 floatfix
]{revtex4-1}

\usepackage[english]{babel}
\usepackage{graphicx}
\usepackage{upgreek}
\usepackage{bm}
\usepackage{textcomp}
\begin{document}

\title{Probing the Pinning Strength of
Magnetic Vortex Cores with sub-nm Resolution
}
\author{Christian Holl}
\email{holl@physik.rwth-aachen.de}
\author{Marvin Knol}
\author{Marco Pratzer}
\affiliation{II. Institute of Physics B and JARA-FIT, RWTH Aachen University, D-52074 Aachen, Germany}
\author{Jonathan Chico}
\author{Imara Lima Fernandes}
\author{Samir Lounis}
\affiliation{Peter Gr\"unberg Institut and Institute for Advanced Simulation, Forschungszentrum J\"ulich and JARA, 52425, J\"ulich, Germany}
\author{Markus Morgenstern}
\affiliation{II. Institute of Physics B and JARA-FIT, RWTH Aachen University, D-52074 Aachen, Germany}
\date{\today}

%\pacs{75.70.Kw}
% 75.70.-i	Magnetic properties of thin films, surfaces, and interfaces
% 75.70.Ak	Magnetic properties of monolayers and thin films
% 75.70.Kw	Domain structure (including magnetic bubbles and vortices)
% 75.75.Fk	Domain structures in nanoparticles
%\keywords{scanning tunneling microscopy, scanning tunneling spectroscopy, magnetic vortex core, magnetic pinning, surface adsorbates}

\maketitle

% ABSTRACT
\textbf{
Topological magnetic textures such as vortex cores or skyrmions are key candidates for non-volatile information processing \cite{Parkin2008, Jung2012, Fert2013, Fert2017}.
This exploits the texture movement by current pulses that is typically opposed by pinning \cite{Jiang2015, Woo2016}. A detailed understanding of pinning is hence crucial with previous experiments being either limited in terms of controlled magnetic texture positioning \cite{Hanneken2016, Attan2001} or in terms of spatial resolution \cite{Cowburn1998, Burgess2013, Holleitner2004, Rahm2004, Klui2003,Badea2016, Basith2012, compton2006}.
Here, we use spin-polarized scanning tunneling microscopy to track a magnetic vortex core that is deliberately moved by a 3D magnetic field.
The core covering ${\sim}10^{4}$ Fe-atoms
gets pinned by defects that are only  a few nm apart.
Reproducing the vortex path via parameter fit, we deduce the pinning potential of the defects as a mexican hat with short-range repulsive and long-range attractive part. By comparison with micromagnetic simulations, the attractive part is attributed to a local suppression of exchange interaction.
The novel approach to deduce defect induced pinning potentials on the sub-nm scale is transferable to other non-collinear spin textures eventually enabling an atomic scale design of defect configurations, e.g., for reliable read-out in race-track type devices \cite{CastellQueralt2019,LimaFernandes2018}.}

% LITERATURE
The intentional pinning of non-collinear magnetic textures by designed defects could enable reliable positioning of domain walls, skyrmions or vortices in future devices such as race-track memories \cite{Parkin2008,Fert2013,Fert2017}.
Magnetic textures are typically pinned in areas of altered magnetization, exchange interaction, anisotropy, or Dzyaloshinskii-Moriya interaction \cite{Paul1982, LimaFernandes2018, Stosic2017, Mller2015}.
Previous experiments probed domain walls \cite{Cowburn1998,Holleitner2004, Basith2012}, magnetic vortices \cite{Burgess2013, Rahm2004, Badea2016, compton2006, Vansteenkiste2008} or skyrmions \cite{Hanneken2016}  embedded in stripes \cite{Holleitner2004, Basith2012}, rings \cite{Klui2003}, islands \cite{Burgess2013, Rahm2004, Badea2016, compton2006, Vansteenkiste2008} or thin films \cite{Cowburn1998, Attan2001, Hanneken2016}
revealing pinning at large, artificial structures (size: $10-100$\,nm) such as notches \cite{Klui2003}, holes \cite{Vansteenkiste2008,Rahm2004} or locally thinned areas of the film \cite{Holleitner2004,Burgess2013} as well as at intrinsic irregularities, e.g., due to surface roughness \cite{Chen2012} or dislocations \cite{Attan2001}. Recently, the pinning of skyrmions at single magnetic impurities has also been probed, but without exerting controlled forces \cite{Hanneken2016}.
Hence, so far experiments were not able to deduce the pinning potential of point defects with the required sub-nm spatial resolution.

% OUR WORK SUMMARY
Here, we employ spin polarized scanning tunneling microscopy (STM) to study pinning of the curled magnetization of a magnetic vortex core \cite{Shinjo2000, wachowiak2002}.
We apply well defined lateral forces by in-plane magnetic fields $\bm{B}_\parallel$ and independently tune the size of the vortex core by an out-of-plane field $B_\perp$.
The latter enables control on the non-collinearity of the core magnetization and, hence, on the strength of exchange energy density $u_{\rm{exch}}$ in the core.
Surprisingly, we find that a vortex core with diameter 3.8\,nm and depth 10\,nm ($\sim$ $10^{4}$ Fe-atoms) jumps between defects only a few nm apart.
The exact pinning position is deduced by measuring topography and core magnetization simultaneously, revealing an eccentric core pinning $\sim$2\,nm away from the next adsorbate.
We reproduce the measured core path along several defects via superposing pinning potentials, each consisting of an attractive part with amplitude 200\,meV originating from quenched $u_{\rm{exch}}$ and an even stronger repulsive part of unknown origin.

% FIGURE 1
\begin{figure}
\includegraphics[width=80mm]{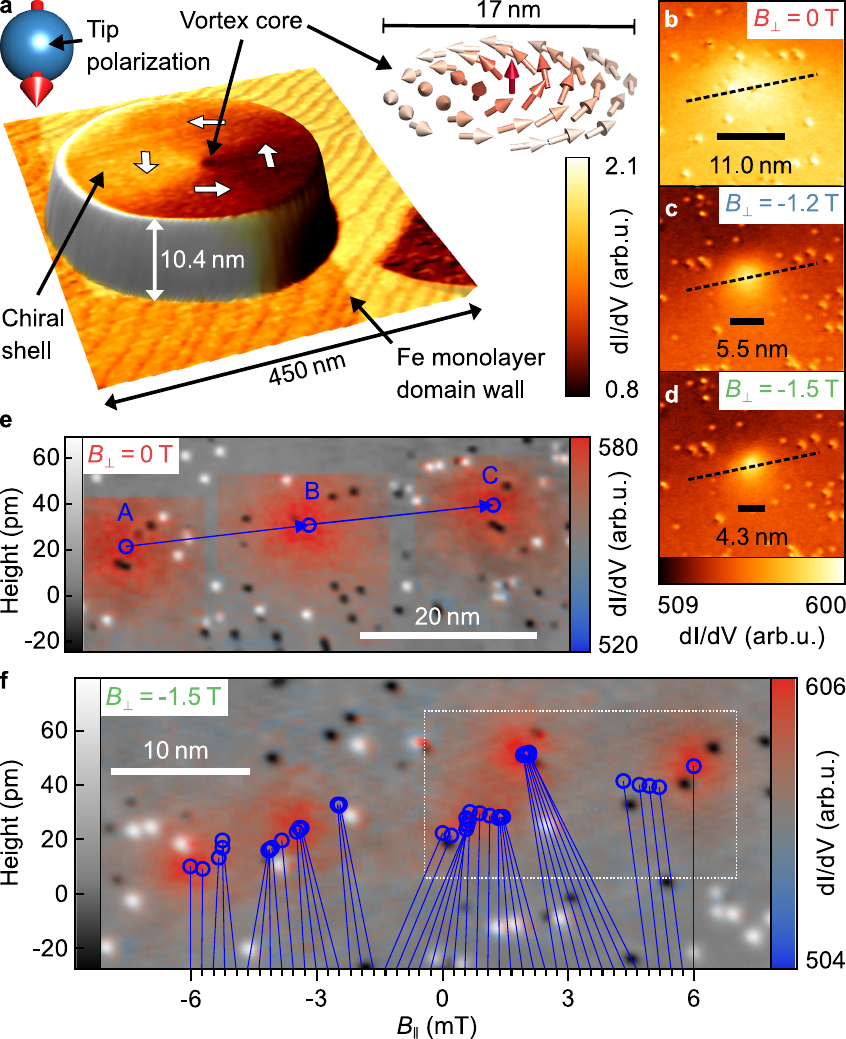}
\caption{\label{fig1}
\textbf{Vortex core trajectories.}
\textbf{a}, Superposition of STM topography (3D representation) and simultaneously acquired spin-polarized d$I$/d$V$ map (color) for an Fe island on W(110), $V = -450$\,mV, $I = 0.5$\,nA. Insets: sketch of deduced tip magnetization vector (left) and spin configuration of the vortex core (right).
\textbf{b-d}, d$I$/d$V$-images of vortex core at identical contrast and different $B_\perp$. The labeled scale bars show FWHM of $m_z$ extracted by core fitting (methods).
\textbf{e}, Superposition of topography (brightness) and three semi-transparent d$I$/d$V$ maps of vortex core (color) after subtracting the signal related to in-plane magnetization (supplement S3) for $\bm{B}_\parallel^{\rm{A}}=(20.5,-11.5)$\,mT, $\bm{B}_\parallel^{\rm{B}}=(16,0)$\,mT, and $\bm{B}_\parallel^{\rm{C}}=(11.5,11.5)$\,mT at $B_\perp=0$\,T. Blue vectors connect the deduced vortex core centers (circles) showcasing the linear core motion.
\textbf{f}, Topography overlaid with vortex core center positions (blue circles) and five selected $dI/dV$ maps (color) for 44 equidistant $\bm{B}_\parallel$ steps with $\Delta \bm{B}_\parallel=(136,-227)$\,$\upmu$T at $B_\perp=-1.5$\,T.
The core center positions are connected to the corresponding $B_\parallel$ (lower axis) by lines.
The d$I$/d$V$ maps (in-plane magnetization subtracted) correspond to $B_\parallel=$ -6\,mT, -3\,mT, 0\,mT, 3\,mT, 6\,mT, $V = -2$\,V, $I = 1$\,nA. A  video of the vortex motion including all 45 $dI/dV$ images is available in the supplementary movies.
The island size is $255\times165\times10$\,nm$^3$ in b-f and $292\times210\times10.4$\,nm$^3$ in a.
}
\end{figure}

% SAMPLE
Atomically flat, elliptical Fe islands with vortex configuration are prepared by molecular beam epitaxy on W(110) (methods) \cite{wachowiak2002, Bode2004}.
STM with antiferromagnetic Cr tips \cite{LiBassi2007} records the  topography and the spin polarized differential conductance d$I$/d$V$ simultaneously (methods).
Figure \ref{fig1}a displays an overlay of these signals for a typical island. The vortex core appears as a dark spot due to the  spin-polarized d$I$/d$V$ contribution proportional to the dot product of tip and sample magnetization vectors \cite{Bode2003}. Defects are visible (Fig.\ \ref{fig1}b-d) via
the non-magnetic part of d$I$/d$V$ probing the local density of states \cite{Bode2003}.

% SQUEEZING CORE
A 3D vector magnet provides $B_\perp$ and $\bm{B}_\parallel=(B_x,B_y)$ \cite{mashoff2009}, hence, tunes the vortex core size and its position, respectively \cite{wachowiak2002}.
Figure \ref{fig1}b-d show d$I$/d$V$-images of the core at increasing $B_\perp $ opposing the core magnetization. The magnetization can be represented by the normalized out-of plane contribution $m_z$.  The Zeeman energy increases and the core diameter shrinks with full width at half maximum (FWHM) of the $m_z$ distribution of $11.0\pm0.1$\,nm (0\,T),  $5.48\pm0.05$\,nm (-1.2\,T) and $4.34\pm0.04$\,nm (-1.5\,T) \cite{wachowiak2002}. This is
reproduced by micromagnetic simulations (supplement S2) implying a large modification of $u_{\rm{exch}}$ at the core center due  to increasing spin canting: 18\,meV/nm$^3$ (0\,T),  95\,meV/nm$^3$ (-1.2\,T), 180\,meV/nm$^3$ (-1.5\,T).
The $u_{\rm{exch}}$ tuning enables varying the vortex-defect-interaction for all defects that modify exchange interaction.

% SHIFTING CORE
Figure \ref{fig1}e reveals the presence of, at least, two types of defects.
The 15\,pm deep depressions are presumably oxygen adsorbates as remainders from the sample preparation. The 40\,pm high protrusions are Cr atoms originating from tip preparation by voltage pulses.
To study the interaction between the vortex core and these defects, we use $\bm{B}_\parallel$ to exert a lateral force on the vortex that shifts the core towards a target position. We monitor the deviation from the target due to defect pinning.
For a defect-free magnetic cylinder, the core position $\bm{r} = (x,y)$ with respect to the island center is adequately described by the rigid vortex model \cite{Guslienko2001}. It minimizes the potential $E(\bm{r}, \bm{B}_\parallel) = \frac{1}{2} k (x^2+y^2)-k\chi_{\rm{free}} (B_y x+B_x y)$ leading to ${\bm r}(\bm{B}_\parallel)= (\chi_{\rm{free}}B_y,\chi_{\rm{free}}B_x)$ \cite{Badea2016}, i.e., the displacement is proportional to $B_\parallel$ with displacement rate $\chi_{\rm{free}}$.
Albeit elliptic islands lead to a directional dependence of $k$ and $\chi_{\rm{free}}$, the core displacement remains largely proportional to $B_\parallel$   (supplement S4).
With additional defects, the potential changes leading to deviations from the regular displacement along a straight path.

% FIGURE 2
\begin{figure*}
\includegraphics[width=175mm]{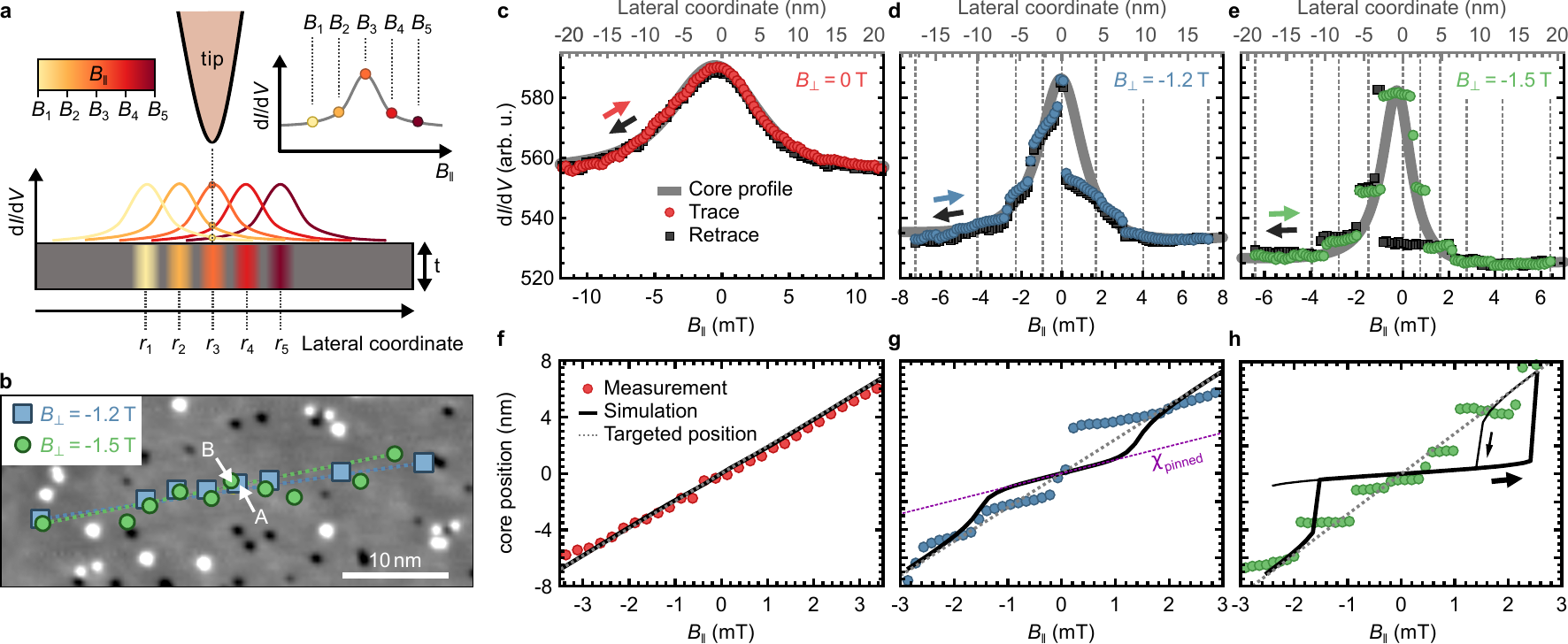}
\caption{\label{fig2}
\textbf{Mapping the strength of vortex core pinning.}
\textbf{a}, Measurement scheme: $\bm{B}_\parallel$ is stepped equidistantly ($B_1$ to $B_5$) to move the vortex core from $r_1$ to $r_5$, while d$I$/d$V$ is recorded at fixed tip position. The resulting d$I$/d$V$($B_\parallel$) displays the core shape in case of a constant core displacement rate.
\textbf{b}, STM topography overlaid with vortex core center positions at two different $B_\perp$ for the $B_\parallel$ highlighted by grey dashed lines in d and e. The positions are deduced from d$I$/d$V$ images at the corresponding $\bm{B}_\parallel$. Dotted lines connect start and end point illustrating the target paths.
\textbf{c,(d,e)}, d$I$/d$V$ recorded at the tip position marked by ``B'' in Fig. \ref{fig1}e (``A'' in b, ``B'' in b)  while sweeping $B_\parallel$ at $B_\perp = 0$\,T ( -1.2\,T, -1.5\,T). The $B_\parallel$ sweep moves the vortex core from ``A'' to ``C'' in Fig. 1e (leftmost to rightmost square or circle in b). The real space d$I$/d$V$ profile recorded along the dashed line in Fig. \ref{fig1}b (\ref{fig1}c, \ref{fig1}d) is plotted in gray. The upper and lower axis are linked by the measured average displacement rate, i.e., $(r_{\rm{C}}-r_{\rm{A}})/(B_{\parallel,\rm{C}} -B_{\parallel,\rm{A}})$  for c and, respectively, for d,e.
\textbf{f-h}, Deduced core positions from c-e assuming a rigid vortex core profile and a straight path (symbols). Solid black lines are micromagnetically simulated core positions for an Fe cylinder (diameter: 280\,nm, thickness: 10\,nm) with a single pinning site  exhibiting $A_{\rm ex} =0$ for a volume of $1.1\times1.1\times0.5$\,nm$^3$ at the surface center. Dotted gray lines show simulated displacement without pinning center. The violet line in g marks the displacement at rate $\chi_{\rm{pinned}}$.
}
\end{figure*}

% REAL SPACE PATH WITH/WITHOUT FIELD
Figure \ref{fig1}e shows the vortex core at three equidistant $\bm{B}_\parallel$ for $B_\perp=0$\,T .
The resulting two displacement vectors exhibit equal lengths $\Delta r = 21.5\pm0.2$\,nm implying a constant displacement rate $\chi(0$\,$T) = 1.74$\,nm/mT as corroborated in Fig.
\ref{fig2}f.
In contrast, Fig. \ref{fig1}f shows irregular vortex core motion for $B_\perp=-1.5$\,T and 45 equidistant $\bm{B}_\parallel$. The core positions are neither equidistant nor along a straight path, but cluster in the vicinity of defects indicating attractive pinning of the core.
Remarkably, a vortex core containing $\sim10^{4}$ Fe atoms (diameter: 3.8\,nm, depth: 10\,nm) is pinned close to a single adsorbate.
Bending of the core in depth direction is relatively small as verified by micromagnetic simulations (supplement S5).

% CORE TRACES
The pinning naturally reduces the displacement rate $\chi$. %$\chi_{\rm{pinned}}(-1.5$\,T$) < \chi_{\rm{free}}(-1.5$\,T).
To determine the resulting $\chi_{\rm{pinned}}(B_\perp)$, a second type of experiment is performed. While the vortex core is displaced by 99 equidistant $ \bm{B}_\parallel$, d$I$/d$V$ is measured at fixed tip position (Fig. \ref{fig2}a).
%These positions are marked B in Fig. \ref{fig1}e and A/B in Fig. \ref{fig2}b.
 We target for the identical defect-free path of length $\sim$35\,nm along several defects for different $B_\perp$ (Fig. \ref{fig2}b).
For $B_\perp=0$\,T, the resulting d$I$/d$V(B_\parallel)$ features an identical shape as the core shape probed by d$I$/d$V ({\bm r})$ in real space at constant $\bm{B}_\parallel$ (gray solid line), i.e.,  d$I$/d$V$ is the same for a tip scanning across a fixed vortex core and for a vortex core scanned below a fixed tip by $\bm{B}_\parallel$.
This confirms, that the large core at $B_\perp = 0$\,T barely interacts with the defects.
In contrast, the datasets at $B_\perp = -1.2$\,T and -1.5\,T show sudden jumps not appearing in the real space data (Fig. \ref{fig2}d,e). They split the curve into segments of reduced slope $\chi_{\rm{pinned}}$ due to core pinning.
The transitions between the segments correspond to jumps between different pinning sites.

% REAL SPACE CONVERSION
To compare these data with theory, we firstly establish a link between the measured d$I$/d$V(B_\parallel)$ and the core displacement (methods).
The conversion uses the real space d$I$/d$V$($\bm{r}$),
implicitly assuming an immutable core profile and a straight core path.
The core shape indeed exhibits negligible FWHM changes by less than $\pm5\%$ along the path (supplement S6).
The motion is not straight (Fig. \ref{fig2}b), but the relatively small excursions imply an error of $\chi_{\rm{pinned}}$ by only 5\,\% (0.3\,\%) at $B_\perp= -1.5$\,T ($-1.2$\,T) (supplement S6).
Figure \ref{fig2}f-h display the converted data. For $B_\perp = $0\,T, we find one constant slope $\chi = \chi_{\rm{free}}(0$\,T$) = 1.8 \pm 0.1$\,nm/mT,
% 1.77$\,nm/mT. I would rather like to use the value of a linear regression.
while, for $B_\perp = -1.2$\,T (-1.5\,T), segments with average slope $\chi_{\rm{pinned}}(-1.2$\,T$) = 1.0 \pm 0.1$\,nm/mT ($\chi_{\rm{pinned}}(-1.5$\,T$) = 0.1 \pm 0.1$\,nm/mT) are interrupted by jumps. A small segment with even negative slope appears (Fig. \ref{fig2}h, $B_{\parallel} = 1-2$\,mT) likely originating from a larger sidewards excursion of the core.
We deduce a large tuning of the displacement rate ratio  $\chi_{\rm{pinned}}/ \chi_{\rm{free}} = 100$\,\%, 42\,\%, and 3\,\% at $B_\perp = 0$\,T, -1.2\,T, and -1.5\,T, respectively.

% FIGURE 3
\begin{figure*}
\includegraphics[width=175mm]{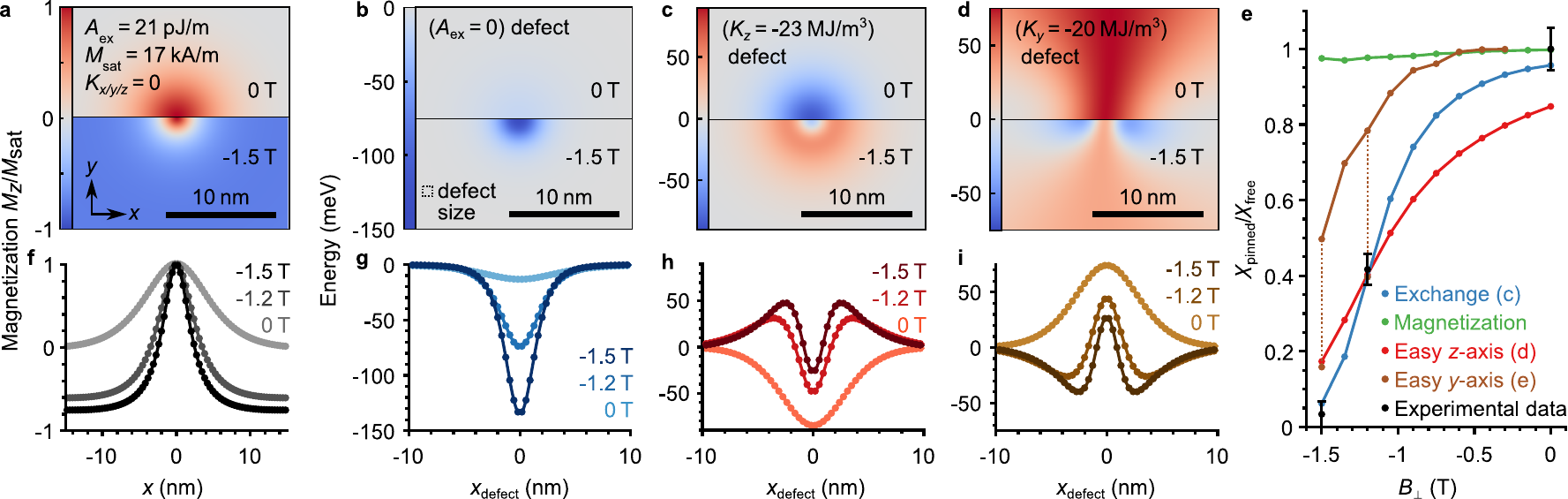}
\caption{\label{fig3}
\textbf{Micromagnetic simulation of pinning potentials.}
\textbf{a}, Scaled, perpendicular magnetization $m_z = M_z/M_{\rm{sat}}$ of a simulated vortex core in a disk of height 10\,nm and diameter 280\,nm at $B_\perp = 0$\,T (upper half) and $B_\perp = -1.5$\,T (lower half) with magnetic parameters indicated.
\textbf{b-d}, Defect potentials for vortex core in meV at different types of magnetic defects, where only the marked parameters are changed with respect to (a) within a central area at the surface of $1.1\times1.1\times0.5$\,nm$^3$ ($3\times3\times1$ cells). The display type is as in a. The spatial dependency of the vortex energy is simulated by scanning the defect  through the vortex core (methods).
%Zero energy corresponds to the value at $x_{\rm{defect}} = -23$\,m for each image.
%The altered magnetic parameters within the defect are indicated in the upper left.
\textbf{f-i}, Profile lines through the middle of a-d (from left to right) covering the 0\,T and the -1.5\,T area separately.
An additional profile calculated for $B_\perp = -1.2$\,T is added.
\textbf{e},  Simulated displacement rate ratio $\chi_{\rm{pinned}}/ \chi_{\rm{free}}$ for the vortex core being trapped in the minima of the potentials shown in g-i. For the $A_{\rm{ex}}$ defect, $A_{\rm{ex}}=0$ is used and the defect size is adapted to fit the experimental data. For the $K_y$ and $K_z$ defects, we kept the defect size, while $K_y$ and $K_z$ are changed to fit the experimental data as good as possible. For the $K_y$ defect, we show two values for the optimized $K_y=300$\,MJ/m$^3$ and the full line with realistic $K_y=20$\,MJ/m$^3$ connected by dotted lines to the optimized points. For the $M_{\rm{sat}}$ defect, we use $M_{\rm{sat}}=0$ within the same defect volume. $B_\perp$ areas providing purely repulsive vortex core potentials are excluded.
Experimental data points are deduced from the average slope of the segments such as in Fig. \ref{fig2}f-h with statistical error bars.
}
\end{figure*}

To reproduce this, we conduct micromagnetic simulations of an Fe cylinder (diameter: 280\,nm, height: 10\,nm) with exchange stiffness $A_{\rm{ex}}$, saturation magnetization $M_{\rm{sat}}$, and uniaxial anisotropy $K_{x/y/z}$ known from previous experiments \cite{wachowiak2002}. The pinning site is modeled by suppressing $A_{\rm{ex}}$, respectively $u_{\rm{exch}}$, within $1.1\times1.1\times0.5$\,nm$^3$.
This defect is moved laterally through the vortex in the island center emulating the vortex movement through the defect by equidistant ${\bm B}_\parallel$ (methods).
The deduced core path as function of $B_\parallel$ is plotted as solid lines in Fig. \ref{fig2}f-h. It reveals slopes $\chi_{\rm{pinned}}$ around $B_\parallel = 0$ very close to the segment slopes of the experimental data, i.e., we find theoretical
$\chi_{\rm{pinned}}/ \chi_{\rm{free}} = 96$\,\%, 40\,\%, and 6\,\% for $B_\perp = 0$\,T, -1.2\,T, and -1.5\,T, respectively. This strongly suggests that quenching of $A_{\rm{ex}}$ is the origin of pinning.

To corroborate this conjecture, simulations are pursued for defects with changed $K_{y}$, $K_{z}$ and $M_{\rm{sat}}$. Figure \ref{fig3}b-d and g-i show resulting pinning potentials deduced from the energy of the vortex core at the corresponding positions (methods). The defect with quenched $A_{\rm{ex}}$ features a purely attractive potential with  an order of magnitude  variation in amplitude by $B_\perp$ (Fig. \ref{fig3}g). The defects with changed anisotropy show more complex potentials with amplitudes that are less dependent on $B_\perp$.

For each kind of defect, we simulated the displacement rate $\chi_{\rm{pinned}}$ around the potential minimum and compared $\chi_{\rm{pinned}}/ \chi_{\rm{free}}(B_\perp)$ with experimental values  (Fig. \ref{fig3}e).
The measured trend is quantitatively reproduced for a defect with quenched $A_{\rm{ex}}$, but not for the other types.
Using $K_z$ and $K_y$ as unrestricted fit parameters,  $\chi_{\rm{pinned}}/ \chi_{\rm{free}}$ can, at most, be reproduced for one of the three $B_\perp$ within error bars. The optimal fitting, moreover, leads to unrealistically large cumulative anisotropies for a single adsorbate: $V_{\rm{defect}}\cdot K_z = 86$\,meV, $V_{\rm{defect}}\cdot K_y=1.1$\,eV (supplement S10). Quenched magnetization does barely pin the core at all implying that quenched $A_{\rm{ex}}$ is indeed the main origin of pinning.

% FIGURE 4
\begin{figure*}
\includegraphics[width=175mm]{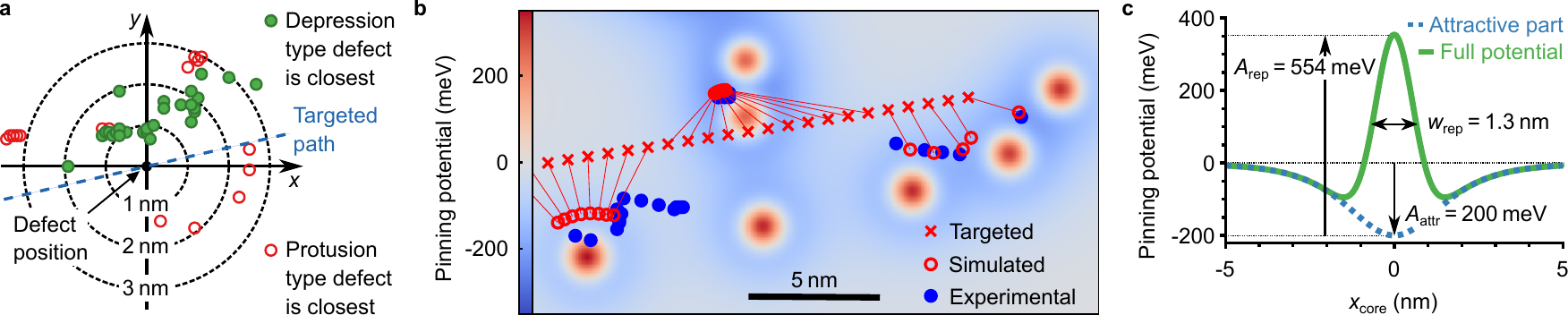}
\caption{\label{fig4}
\textbf{Extracting the defect interaction potential.}
\textbf{a}, Vortex core positions (symbols) with respect to the position of the closest adsorbate, differently colored for the two types of adsorbates. Targeted path (dashed line) reveals that pinning is mostly offset perpendicular to the target path.
\textbf{b}, Simulation of the vortex core path (red circles) within the displayed interaction potential (color map) that superposes the same pinning potential as displayed in c centered at each adsorbate. Adsorbate positions are taken from topography (Fig. \ref{fig1}f, white box). Experimental core positions (blue points, Fig. \ref{fig1}f) and target core positions (red crosses) are added.
\textbf{c}, The optimized, axially symmetric single defect potential consisting of an attractive part due to quenched exchange energy (Fig. \ref{fig3}g) and a repelling Gaussian part.  The three relevant fit parameters are marked.
}
\end{figure*}

% FITTING OF THE REPELLING POTENTIAL
%Surprisingly, albeit the pinning strength in Fig. \ref{fig3}b,g is strongest in the vortex core center, the pinning positions of the cores are offset from the adsorbates (Fig. \ref{fig1}f).
Next, we evaluate the precise pinning position of the vortex core center with respect to the closest adsorbate (Fig. \ref{fig4}a). They cluster at 1-2\,nm away from the adsorbate indicating an additional repulsion.
Moreover, the offset is mostly directed perpendicular to the target path as expected for an isotropic potential preferentially attracting an object along a line perpendicular to its target path.

To estimate the repelling part of the potential, we employed a fit of 24 subsequent experimental pinning positions (blue dots, Fig. \ref{fig4}b) by adapting three parameters for an identical potential centered at each adsorbate, namely a scaling factor for the axially symmetric ($A_{\rm{ex}}=0$)-potential (Fig. \ref{fig3}g, $B_\perp=-1.5$\,T)
as well as height and FWHM of an axially symmetric Gaussian repelling part (Fig. \ref{fig4}c).
The energetic cost of moving the core from the target path towards pinning is firstly calculated without defects via micromagnetic simulations of the vortex energy required to force the core away from its target path. Subsequently, this energy is combined with the pinning potentials yielding the minimum energy position (methods).
Figure \ref{fig4}b shows rather good agreement of resulting optimized path (red circles) and measured core positions (blue circles) employing the defect potential of Fig. \ref{fig4}c.

It is a mexican hat with minima located 1.5\,nm away from the center as expected from the pinning positions (Fig. \ref{fig4}a). The mexican hat also reproduces the queuing of the core in front of the double defect located above the target path (Fig. \ref{fig4}b).
This queuing is markedly different from the slow motion during pinning at a single defect.
It cannot be reproduced by overlapping two ($A_{\rm{ex}}=0$)-potentials with arbitrary independent positions and, hence, corroborates the mexican hat shape.
%
%{\color{red}Samir: Would adding the following help? We note that potentials with such a shape can be triggered by the subtle balance of oscillatory magnetic interactions as predicted by ab-initio simulations for small magnetic skyrmions interacting with single transition metal impurities.\cite{LimaFernandes2018}}
%
Naturally, the ($A_{\rm{ex}}=0$)-part of Fig. \ref{fig3}g has to be rescaled to compensate for the repelling part, i.e. the ($A_{\rm{ex}}=0$)-defect has to be slightly enlarged.

% AB INITIO
We were not able to pinpoint the origin of the repulsive part. Since it is smaller than the vortex core, it cannot be reproduced by simply changing parameters constantly within a certain area. We refrained from optimizing more complex defect structures avoiding the increasing parameter space.

Employing simulations based on density functional theory (DFT), we investigated the impact of single Cr and O adatoms on the magnetic properties of Fe(110). We find remarkably strong changes of the pairwise magnetic exchange interactions $J_{ij}$ ($i$, $j$: atomic sites) affecting up to $~70$ neighboring Fe atoms (supplement S10).
The summed up change
is $\sim 200$\,meV consisting
of similar amounts of weakening and strengthening of $J_{ij}$ due to the oscillatory behavior of the interactions as function of distance. Hence,  the sum of changes of $|J_{ij}|$ amounts to $2.5$\,eV. However, if the vortex core texture is not changed by the defect as implied by the barely changing spin contrast in STM (supplement S6), the amplitude of the core-adsorbate interaction amounts to only $10-15$\,meV (supplement S10).
Thus, while the DFT results reveal that single Cr or O adsorbates influence the core path on the 0.5\,nm scale (supplement S10), they do not explain the experiments quantitatively.
%Hence, the single adsorbate as described by DFT cannot explain our results, albeit it would influence the path of the core on the 0.5\,nm scale (supplement).
We speculate that the adsorbate structure is either different than anticipated or that the adsorbate is accompanied by particular strain fields below the surface accounting for the missing energy.

% PERSPECTIVE
Our novel method provides the first quantitative handle on pinning energies of magnetic textures at the sub-nm scale.
In principle, it can be applied to different kinds of deliberately placed defects on different types of magnetic islands featuring vortices.
It can also be used for other non-collinear textures such as skyrmions or transverse domain walls anticipated to be used in racetrack memories \cite{Parkin2008, Fert2013, Fert2017}.
Both have been imaged by spin polarized STM \cite{Romming636, Bode2003, Hanneken2016}. For skyrmions, additionally the spin canting and, hence, $u_{\rm{exch}}$ can be tuned by $B_\perp$ \cite{Romming2015}.
Forces on domain walls can be exerted by ${\bm B}_\parallel$, \cite{Ono1999,Klui2003}, while skyrmions can be moved by electric currents \cite{Jiang2015,Woo2016}, for which respective forces are deduced by combining micromagnetic simulations and an analytic description via the Thiele equation \cite{Sampaio2013}.
This would enable experimental probing of the theoretically predicted skyrmion-defect interaction strenghts \cite{LimaFernandes2018, Lin2013, Mller2015, Stosic2017}.
Eventually, our method could provide tailoring rules for defect induced guiding of magnetic textures in racetrack memories \cite{CastellQueralt2019,LimaFernandes2018}.

\begin{acknowledgments}
We gratefully acknowledge insightful discussions with H.-J. Elmers, S. Bl\"ugel, A. Schlenhoff, M. Liebmann and financial support of the German Science Foundation (DFG) via PR 1098/1-1 and of the European Research Council (ERC) under the European Union’s Horizon 2020 research and innovation program (ERC-consolidator grant 681405 — DYNASORE). We are grateful for the computing time granted by the JARA-HPC Vergabegremium and VSR commission on the supercomputer JURECA at Forschungszentrum J\"ulich and at the RWTH Aachen supercomputer.
\end{acknowledgments}

%%%%%%%%%%%%%%%%%%%%%%%%%%%%%%%%%%%%%%%%%%%%%%%%
% METHODS
%%%%%%%%%%%%%%%%%%%%%%%%%%%%%%%%%%%%%%%%%%%%%%%%

\section{Methods}
\textbf{Preparation.}
A W(110) crystal (surface orientation better than 0.1\textdegree) is cleaned in ultra high vacuum (UHV) (base pressure: 10$^{-10}$\,mbar) by repeated cycles of annealing in oxygen atmosphere (partial pressure: 10$^{-7}$\,mbar) at 1400\textdegree C for 10\,min and subsequent flashing to 2200\textdegree C for 10\,s. Afterwards, ten pseudomorphic monolayers of Fe are deposited at room temperature by electron beam evaporation from an Fe rod (purity 99.99+\%). The sample is then annealed at 710\textdegree C for 20\,min leading to the formation of Fe islands such as in Fig. \ref{fig1}a on top of an Fe wetting layer \cite{Bode2004}.

\textbf{Spin polarized STM.}
The tunneling tip is fabricated from a $0.5\times0.5$\,mm$^2$ beam of polycrystalline, antiferromagnetic Cr (purity 99.99+\%). Tip sharpening employs electrochemical etching by a suspended film of 2.5\,M NaOH solution within a PtIr loop that is at potential of 5.5\,V with respect to the tip. Etching is stopped at drop off of the lower beam part via differential current detection. The upper part of the beam is immediately rinsed with DI water and glued onto a custom tip holder. The tip is then loaded into the UHV system and, subsequently, into the STM scan head at 6\,K \cite{mashoff2009}.
The atomic structure of the tip is optimized during tunneling by  voltage pulses (10\,V/30\,ms) between tip and sample until spin contrast is achieved. Voltage $V$ is applied to the sample. The differential conductance d$I$/d$V$ is measured by adding a 50\,mV RMS sinusoidal voltage (1384\,Hz) to the applied DC $V$ and recording the resulting oscillation amplitude of the tunnel current $I$ using a lock-in amplifier. The system enables a 3D magnetic field ${\bm B}=(B_x,B_y,B_\perp)$ with out-of-plane component $B_\perp$ up to 7\,T and simultaneous in-plane part ${\bm B}_\parallel =(B_x,B_y)$ up to 1\,T in each in-plane direction \cite{mashoff2009}.

\textbf{Micromagnetic simulations.}
The program mumax$^3$ \cite{Vansteenkiste2014} is used to simulate relaxed magnetization states of an Fe cylinder of height 10\,nm and diameter 280\,nm with cell size $0.36\times0.36\times0.5$\,nm$^3$. Magnetic parameters are marked in Fig. \ref{fig3}a. Defects are emulated by altered magnetic parameters in  $3\times3\times1$ cells at the top layer. For sweeps of $B_\parallel$ with defect, two approximations are employed in order to reduce computational time. Instead of sweeping $B_{\parallel} = B_{\parallel,\rm{target}}$, we keep $B_{\parallel}=0$\,T and shift the defect through the vortex core  by $-\chi_{\rm{free}}\cdot B_{\parallel,\rm{target}}$ with $-\chi_{\rm{free}}$ deduced from a simulation of the vortex with varying $B_{\parallel}$ but without defects. Second, we crop the simulation area down to $256\times256\times20$ cells via adding the previously calculated demagnetization field of the neglected area manually. This leads to an effective, spatially varying external magnetic field $\bm{B}_{\rm{eff}} ({\bm{r}})=\bm{B}_\perp+\bm{B}_{\rm{demag, exterior}}({\bm{r}})$.
The reasonable validity of these approximations is described in supplement S7. The resulting core center positions ($m_z$ maxima) as a function of defect position are deduced from spline interpolations of $m_z$ in the layer below the defect. This avoids the more ambiguous evaluation of the partially discontinuous $m_z$ within the surface layer in the presence of defects.

\textbf{Vortex core fitting to determine its center position.}
To reproduce the experimental spin polarized $dI/dV$ images and, hence, to deduce the vortex core center positions, vortex magnetization patterns are firstly simulated via mumax$^3$.
The result is then adapted to the experimental $dI/dV$ image at corresponding $B_\perp$. Therefore, the polar and azimuthal angle of the tip magnetization are optimized using the dot product between sample and tip magnetization vector as $dI/dV$ image contrast.
The resulting $dI/dV$ values are additionally offset and scaled to account for the non-spin-polarized d$I$/d$V$ signal and the unknown amplitude of the spin-polarized d$I$/d$V$ signal, respectively.
Moreover, the vortex core center position is optimized in both lateral directions and the calculated image is slightly scaled laterally to account for inaccuracies of the fit (supplement S2).

The seven parameters (2 tip magnetization angles, $dI/dV$ offset, $dI/dV$ scaling factor, $2 \times$ core position, lateral scaling factor) are fitted towards minimum RMS deviation between the simulated and the measured d$I$/d$V$ map. The blue circles in Fig. \ref{fig1}e-f as partly also displayed in Fig.\,\ref{fig4}b and the squares and circles in Fig. \ref{fig2}b are the fitted lateral positions of the vortex core center with each symbol belonging to a fit of one d$I$/d$V$ map.
The fit error in core center position turns out to be $\le \pm 0.05$\,nm. Fit images, residual images and standard deviations for all fit parameters are given in  supplement S3.

For the superposition of topography and sequences of d$I$/d$V$ data (Fig. \ref{fig1}e-f), the in-plane magnetization contribution of the fitted $dI/dV$ image is removed from the measured one and, for Fig. \ref{fig1}f, the resulting image is scaled by a Gaussian envelope function for the sake of visibility.

\textbf{Conversion from d$I$/d$V(B_\parallel)$ to core positions.}
To calculate a vortex core position from a d$I$/d$V$ value measured at fixed tip location $\bm{r}_0$ but varying ${\bm B}_\parallel$, we use the line profile d$I$/d$V(r')$ of the vortex core measured at constant $\bm{B}_\parallel$ (Fig. \ref{fig1}b$-$d).
We first employ the fit procedure as explained in the previous section and then utilize the less noisy profile from the fitted, simulated $dI/dV$ images.
Angle of chosen profile line and lateral shift of the profile line with respect to the core center are selected such that the maximum value in d$I$/d$V(B_\parallel)$ and the d$I$/d$V$ values at maximum and minimum of $B_\parallel$ (Fig. \ref{fig2}c-e) are reproduced by a straight target path (supplement S4).
 The parameter $r'$ is set to zero at  maximum d$I$/d$V$.  Using the resulting d$I$/d$V(r')$, the measured d$I$/d$V(B_\parallel)$ at $\bm{r}_0$ is assigned to a core center position $\bm{r}_0+\bm{u}r'$(d$I$/d$V(B_\parallel)$) with $\bm{u}$ being the unit length vector in the selected profile direction.
Principally, there are two possibilities of $r'$(d$I$/d$V(B_\parallel)$), left and right from the center of the profile line. They are handled such that the core center always moves to the closer of the two $r'$ and continuously across $\bm{r}_0$.

\textbf{Calculating vortex core pinning potentials}
To calculate the pinning potentials as displayed in Fig. \ref{fig3}, the parameters are homogeneously changed within $3\times3\times1$ cells mimicking the defect. Subsequently the defect is moved through the fixed vortex core and the resulting vortex energy is calculated by mumax$^3$. The approximation to move the defect instead of the vortex core is discussed in supplement S7.

\textbf{Simulating the vortex path for multiple defects.}
The vortex core position for an immutable core profile is given by minimizing the potential energy $E_{\rm{pot}}(\bm{r}_{\rm{vortex}}) = E_{\rm{flex}}(\bm{r}_{\rm{vortex}}-\bm{r}_{\rm{target}}) + \sum_{i=1}^N E_{\rm{i,pin}}(\bm{r}_{\rm{vortex}}-\bm{r}_{\rm{i,adsorbate}})$. $E_{\rm{flex}}(\bm{r}_{\rm{vortex}}-\bm{r}_{\rm{target}})$ is the energetic cost to move the vortex away from its target $\bm{r}_{\rm{target}}(B_\parallel)$ in the absence of defects.
It is deduced from a set of mumax$^3$ simulations fixing the vortex core artificially at different $\bm{r}_{\rm{vortex}}$. This employs fixing $m_z$ within
$4\times 4$ cells on the surface located away from $\bm{r}_{\rm{target}}$. The $m_z$ values in that area are set to the values found in the center of the vortex core, if calculated without defects.
The vortex core, consequently, moves to a particular ${\bm r}_{\rm{vortex}}$ with respect to $\bm{r}_{\rm{target}}$. For this position, we calculate the vortex energy. We checked that the area of fixed $m_z$ leads to negligible changes of the vortex energy (supplement S8).
For sake of simplicity, we approximate the resulting $E_{\rm{flex}}(\bm{r}_{\rm{vortex}}-\bm{r}_{\rm{target}})$ by an excellently fitting paraboloid (supplement S8).

The pinning potential of a single adsorbate $E_{\rm{i,pin}}(\bm{r}_{\rm{vortex}})$ is emulated as the sum of a repelling Gaussian and an analytic representation of the attractive part due to a defect with quenched $A_{\rm{ex}}$. This pinning potential eventually reproduces the profile of Fig. \ref{fig4}c by fitting the core path in Fig. \ref{fig4}b. The analytic representation of the attractive part is derived straightforwardly from an analytic part of the core magnetization profile reading $m_z(r) = (1-a)/\cosh(|r|/w)+a$ with fit parameters $a$ and $w$ \cite{3540641084}.
The deduced analytic $u_{\rm{exch}} (r)$ is fitted to the result from mumax$^3$ (Fig. \ref{fig3}g) with respect to $a$ and $w$ exhibiting an RMS deviation of only 0.6\,mV between analytic and micromagnetic representation of $u_{\rm{exch}} (r)$ (supplement S8).

The subsequent fitting of the core path optimizes FWHM and amplitude of the repelling Gaussian as well as a scaling parameter for the  attractive, analytic exchange part (Fig. \ref{fig4}c) towards minimizing the RMS of the distances between calculated and measured core center positions (Fig. \ref{fig4}b).
Additionally, the start and end point of the target path are varied by up to $\pm 3$\,nm during the fit with respect to the observed first and last core positions to account for possible pinning at these sites.

\section{Supplementary information}
\subsection*{S1: Fe island}
\label{sec:island}

\begin{figure}
\includegraphics[width=14cm]{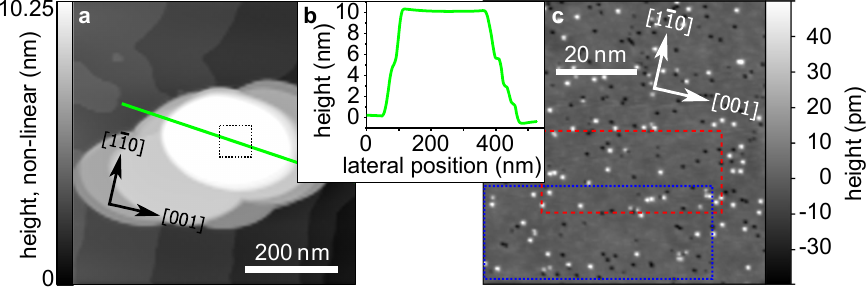}
\caption{\label{figS1}
\textbf{Investigated Fe island.}
\textbf{a}, Topographic image of the island displayed in non linear gray-scale to enhance the visibility of substrate step edges. The island is imaged multiple times due to tip artifacts.
\textbf{b}, Topographic profile along the green path in a. The average island height is 10\,nm.
\textbf{c}, Zoom into the area marked by the dashed box in a showing the adsorbates. Red box depicts the area imaged in  Fig. 1e of the main text, blue box depicts the area imaged in Fig. 1f of the main text \cite{thesis_holl_2018}.
}
\end{figure}

Figure \ref{figS1}a shows an STM image of the Fe island that has been studied in Fig.\,1b-f and Fig.\,2 of the main text. Its size is $255\times165\times10$\,nm$^3$. The crystallographic axes of the substrate as deduced from a low-energy electron diffraction pattern are added. The topographic image suffers from a multi-tip artifact that images the island several times. This does not influence spectroscopic measurements on the topmost imaged surface since the additional tips are a few nanometers away from that surface during its measurement.

\clearpage

%%%%%%%%%%%%%%%%%%%%%%%%%%%%%%%%%%%%%%%%%%%%%%%%%%%%%%%%%%%%%%%%%%%%%%%%%%%%%%%%%%%%

\subsection*{S2: Micromagnetic energy densities of vortex core}
\label{sec:energy density}

\begin{figure}
\includegraphics[width=16.4cm]{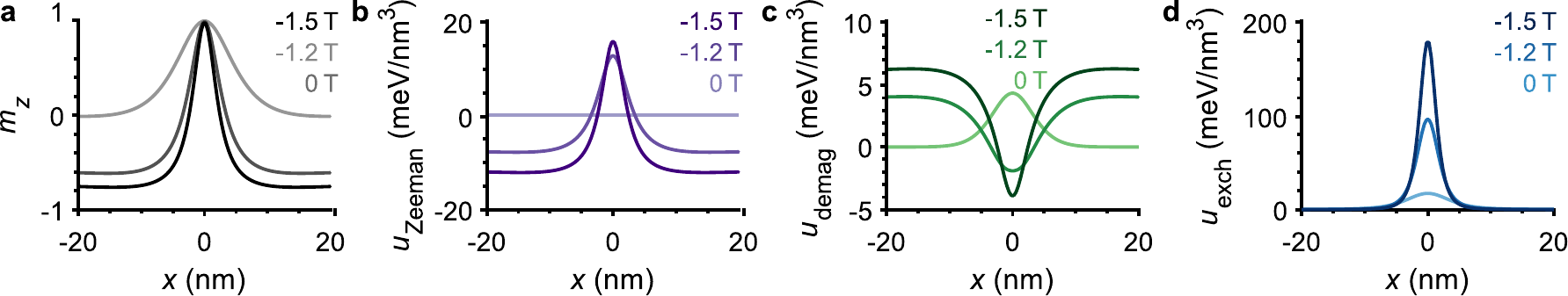}
\caption{\label{figS4}
\textbf{Vortex core energy densities.}
\textbf{a}, Profiles of perpendicular magnetization $m_z = M_z/M_{\rm{sat}}$ of a simulated vortex core in a disk of height 10\,nm and diameter 280\,nm at $B_\perp$ according to legend.
\textbf{b-d}, Profiles of Zeeman energy density, demagnetization energy density, and exchange energy density.
}
\end{figure}

Magnetic vortex patterns are relaxed within the micromagnetic software package mumax$^3$ for a circular Fe island of thickness 10\,nm and diameter 280\,nm at perpendicular fields of $B_\perp = 0$\,T, -1.2\,T, and -1.5\,T. The simulation space is discretized into $768\times768\times1$ cells of size $0.364\times0.364\times10$\,nm$^3$. Magnetic parameters are set to saturation magnetization $M_{\rm{sat}} = 17$\,kA/m, exchange stiffness $A_{\rm{ex}}=21$\,pJ/m, and zero magnetocrystalline anisotropy \cite{wachowiak2002}. Spatially resolved energy densities of the Zeeman term, the demagnetization and the exchange are output by the software after relaxation of the magnetization pattern. Profiles through the vortex center of the cylindrical symmetric energy densities are shown together with profiles of the scaled out-of-plane magnetization $m_z$ in Fig. \ref{figS4}.

The $m_z$ profiles (Fig.\,\ref{figS4}a) largely map the experimentally observed $dI/dV$ images, in particular, if the exact shape of the island is taken into account
(section\,\ref{sec:corefit}). The exchange energy densities (Fig.\,\ref{figS4}d) are
much larger than the other two energy contributions.
They, moreover, vary by approximately an order of magnitude with $B_\perp$, which results in a strong variation of pinning strength with $B_\perp$ for a defect with quenched $A_{\rm{ex}}$, as described in the main text.

\clearpage

%%%%%%%%%%%%%%%%%%%%%%%%%%%%%%%%%%%%%%%%%%%%%%%%%%%%%%%%%%%%%%%%%%%%%%%%%%%%%%%%%%%%

\subsection*{S3: Core fitting procedure}
\label{sec:corefit}

\begin{figure}
\includegraphics[width=16.4cm]{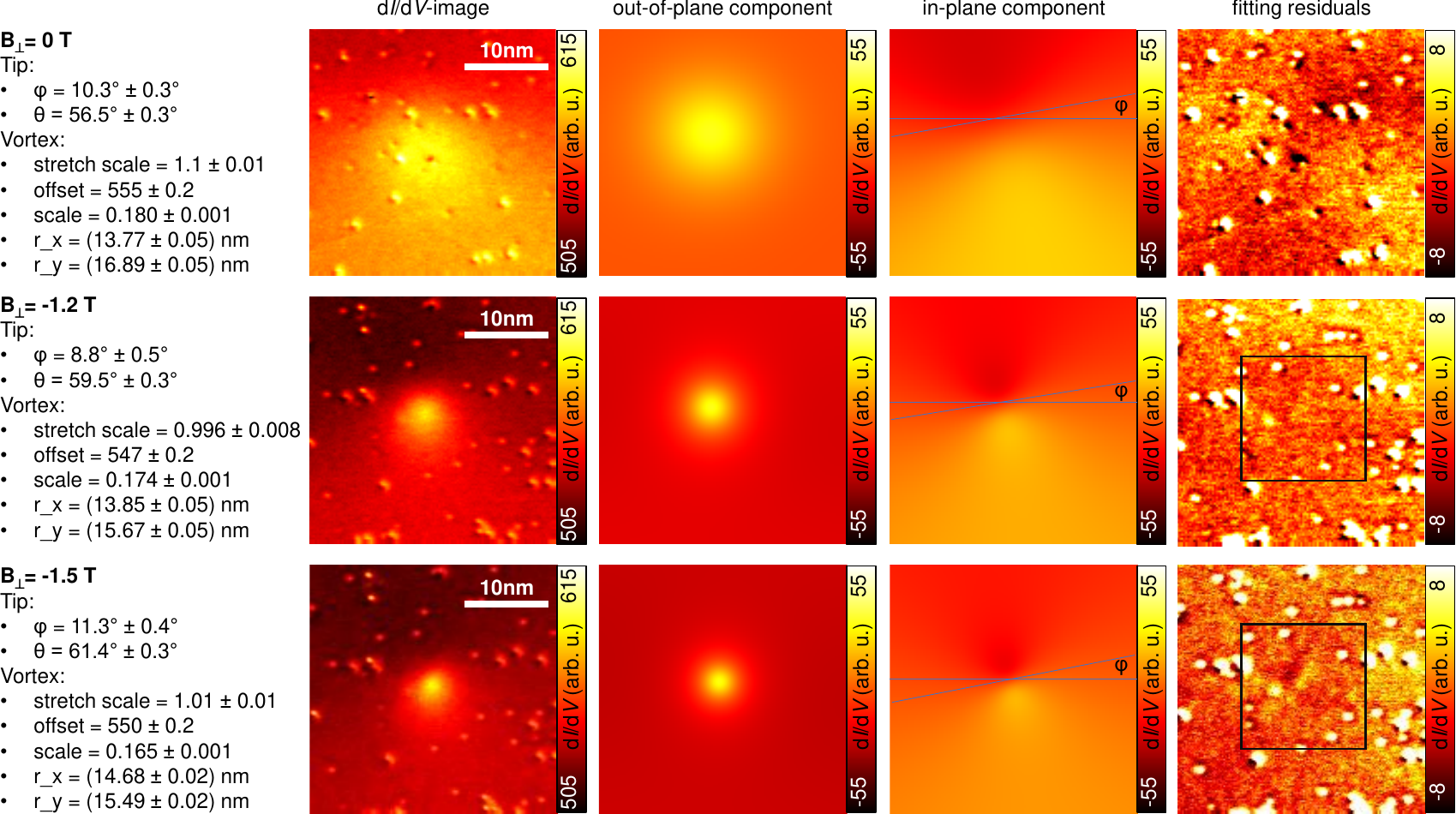}
\caption{\label{figS7}
\textbf{Fitting the $dI/dV$ image of a vortex core by micromagnetically simulated $m_z$ profiles.}
Each row belongs to one $B_\perp$ as marked. The columns show (left to right): Fit parameters, original d$I$/d$V$ images (same as Fig. 1b-d of main text), fitted out-of-plane magnetization component of $dI/dV$, fitted in-plane magnetization component of $dI/dV$, residual image. Note the larger contrast scale of the residual image by about an order of magnitude with respect to the other images.
The fit parameters are azimuthal angle $\phi$ and polar angle $\theta$ of tip magnetization, a lateral scaling factor for the micromagnetically simulated images called stretch factor, a $dI/dV$ offset and $dI/dV$ scaling factor to account for non-spin-polarized d$I$/d$V$ background and magnitude of spin polarized $dI/dV$ signal, respectively \cite{Bode2003}, and the desired core center position $(r_x,r_y)$ \cite{thesis_holl_2018}. The black squares in the two lower right images mark the area where the fit is optimized. The fit angle $\phi$ is indicated in the images of the forth column.
}
\end{figure}

Figure \ref{figS7} visualizes the fitting procedure for $dI/dV$ images of the vortex core as shown in Fig. 1b-d of the main text
 and again in the 2$^{\rm{nd}}$ column of Fig. \ref{figS7}.

The micromagnetic simulations employed for the core fits are conducted for an Fe island with thickness of 10\,nm and lateral shape as determined by STM experimentally. The island is discretized in cells of  $0.359\times0.359\times1$\,nm$^3$. The experimental $dI/dV$ images and the micromagnetically simulated magnetization images are firstly interpolated to the same resolution.
Moreover, defects are removed from the experimental image by a masking procedure prior to the fitting.
Fit parameters are the two angles of the tip magnetization vector, the core position $(r_x,r_y)$, a small lateral scale factor for the simulated images as well as the required offset and scale factor to transfer the dot product of magnetization vectors of tip and sample to the simulated $dI/dV$ value \cite{Bode2003}.

The seven fit parameters are optimized towards minimum RMS deviation between simulated and measured d$I$/d$V$ map employing the MATLAB inbuilt trust-region-reflective least squares algorithm.
At larger $B_\perp$, we only use the displayed black squares in the right column of Fig.\,\ref{figS7} for optimization such that we get more sensitive to the core region.
The quality of the fits is visible in the most right column of Fig. \ref{figS7}
showcasing the residual images that are obtained by subtracting the simulated $dI/dV$ image from the experimental one. Only the adsorbates on the surface are visible with barely any magnetic contrast originating from the vortex core, even at the tenfold increased contrast scale of the residual images with respect to the experimental $dI/dV$ images.
The resulting fit parameters and confidence intervals are given in the left column of Fig.\,\ref{figS7}.

The fit parameters firstly reveal a tip magnetization that slightly cants into the out-of plane direction with increasing $B_\perp$ as expected. Moreover, the stretch scale is very close to one at larger $B_\perp$, while deviating by 10\% at $B_\perp = 0$\,T.
In line, the residual contrast surrounding the core is more pronounced at $B_\perp = 0$\,T, where it features four areas of alternating bright and dark contrast. This is likely caused by the influence of adsorbates on the in-plane magnetization that prohibits a perfect fitting by the micromagnetic vortex simulated without defects. In line, the stretch factor at larger $B_\perp$ also deviates from one by $\sim 10$\%,  if the adapting area is not reduced to the displayed square.
The deduced $dI/dV$ offset and $dI/dV$ scale are very similar for all three $B_\perp$. The obtained large consistency of all fit parameters implies that the fits are reliable, in particular, at larger $B_\perp$, enabling a rather precise determination of the core center position of the vortex.

 Via the extracted angle $\theta$ of tip magnetization, we, moreover, can discriminate the out-of-plane contrast and the in-plane contrast of the $dI/dV$ images as displayed in the third and forth column of Fig. \ref{figS7} for the simulated $dI/dV$ images. The in-plane angle of tip magnetization $\phi$ is additionally marked. The discrimination is used to display an overlap of several vortex cores in one image as in Fig. 1e-f of the main text. To improve the visibility of each core, we subtract the in-plane contrast from the experimental $dI/dV$ images. For Fig. 1f of the main text and the second supplemental movie, we afterwards multiply the remaining out-of-plane contrast including defects by a Gaussian envelope centered at the vortex core center. This makes following the vortex core visually significantly more easy.

\clearpage

%%%%%%%%%%%%%%%%%%%%%%%%%%%%%%%%%%%%%%%%%%%%%%%%%%%%%%%%%%%%%%%%%%%%%%%%%%%%%%%%%%%%

\subsection*{S4: Core movement in elliptic island}
\label{sec:elliptic}

As described in the main text, the lateral core position ${\bm r}$ induced  by ${\bm B}_\parallel = (B_x,B_y)$ follows ${\bm r}(\bm{B}_\parallel)= (\chi_{\rm{free}}B_y,\chi_{\rm{free}}B_x)$ with  displacement rate $\chi_{\rm{free}}$ for a circular magnetic island \cite{Badea2016}.
We assume a similar relation ${\bm r}(\bm{B}_\parallel)= (\chi_{x,\rm{free}}B_y,\chi_{y,\rm{free}}B_x)$ for the investigated elliptical island. This allows us to deduce target positions from ${\bm B}_\parallel$ by a shift ${\bm r} = (\chi_{x,\rm{free}} B_y,\chi_{y,\rm{free}} B_x)$ from the starting point ${\bm r}({\bm 0}$\,T$)={\bm 0}$\,nm.
The validity of this assumption is verified by micromagnetic simulations revealing that a change of  $\bm{B}_\parallel$ by $\Delta \bm{B}_\parallel$ results in nearly identical core shifts $\Delta{\bm r}$ independent of $\bm{B}_{\parallel}$.
We employ a $3 \times 3$ grid of simulations with equidistant $\bm{B}_{\parallel}$ using the experimental island shape with cell size $0.36\times0.36\times10$\,nm$^3$ and magnetic parameters as displayed in Fig.\,3a of the main text.

\begin{figure}
\includegraphics[width=16.4cm]{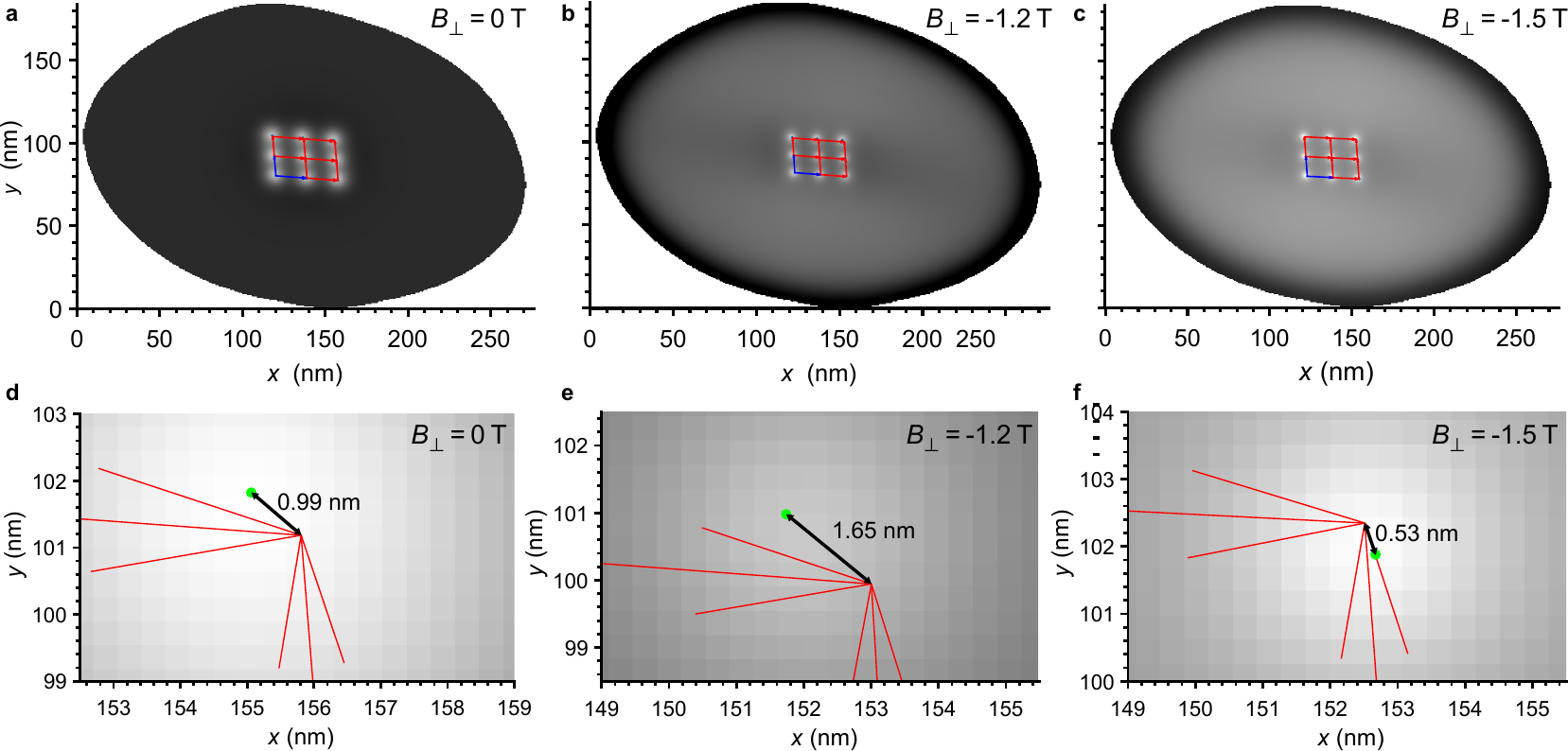}
\caption{\label{figS2}
\textbf{Validity of constant displacement rate by $B_\parallel$ in an elliptical island.}
The program mumax$^3$ is used to simulate the movement of the vortex core within an island featuring the shape of the experimental one.
Nine  positions are targeted using a $3\times 3$ grid of equidistant ($B_x$,$B_y$). \textbf{a}-\textbf{c},
Grayscale plots of cumulative $m_z$ of all nine simulations featuring all nine vortex cores for each $B_\perp$. The contrast is adapted to minimum and maximum of $m_z$ in each image individually. Blue vectors interconnect the lower left vortex core center to its two neighbors. Red vectors result from shifting the blue vectors in order to continue the lattice. \textbf{a}, $B_\perp = 0$\,T, $B_x$ = 12/0/-12\,mT, $B_y$ = -12/0/12\,mT.
\textbf{b},  $B_\perp = -1.2$\,T, $B_x$ = 8/0/-8\,mT, $B_y$ = -8/0/8\,mT.
\textbf{c}, $B_\perp = -1.5$\,T,  $B_x$ = 7/0/-7\,mT, $B_y$ = -7/0/7\,mT. \textbf{d}$-$\textbf{f}, Zoom into the upper right vortex core area of the $3\times 3$ grid in a$-$c, respectively. The green dots mark the simulated vortex core center deduced from the $m_z$ maximum as found by spline interpolation.
Red arrows are the end points of the continuation vectors from a$-$c. The mismatch between the vector addition (red arrows) and the simulated core positions is marked being 1$-$4\,\% of the full distance of movement of 37\,nm.
}
\end{figure}

Fig.\,\ref{figS2}a$-$c show an $m_z$ overlay of the resulting nine vortex cores with centers connected by colored vectors for each of the three experimental $B_\perp$. The bottom left vortex core is used as reference point with two lattice vectors (blue) to its nearest neighbors. These vectors set the displacement rates $\chi_{x,\rm{free}}$ and $\chi_{y,\rm{free}}$.
Assuming constant $\chi_{i,\rm{free}}$, the red vectors mark the lattice continuation that roughly hits the other calculated vortex cores. Zooming into the area of the upper right core (Fig.\,\ref{figS2}d$-$f) reveals a remaining mismatch of $\sim 1$\,nm. This corresponds to a displacement error of $\sim 3$\,\% on the full range of 37\,nm of core movement, directly translating to an error of the anticipated constant $\chi_\mathrm{free}$ in Fig.\,2f$-$h of the main text. Note that the distance of the simulated core movement in Fig.\,\ref{figS2} is identical to the experimental one in Fig.\,2f$-$h of the main text.

\clearpage

%%%%%%%%%%%%%%%%%%%%%%%%%%%%%%%%%%%%%%%%%%%%%%%%%%%%%%%%%%%%%%%%%%%%%%%%%%%%%%%%%%%%

\subsection*{S5: Core bending within Fe island by pinning at the surface}
\label{sec:bending}

Spin polarized STM probes the magnetization of the surface layer that could be distinct from the magnetization in deeper layers. In particular, if the pinning center is at the surface only, the vortex core might bend towards its target position in deeper layers. Here, we show that the resulting vortex core bending is small.

\begin{figure}
\includegraphics[width=15cm]{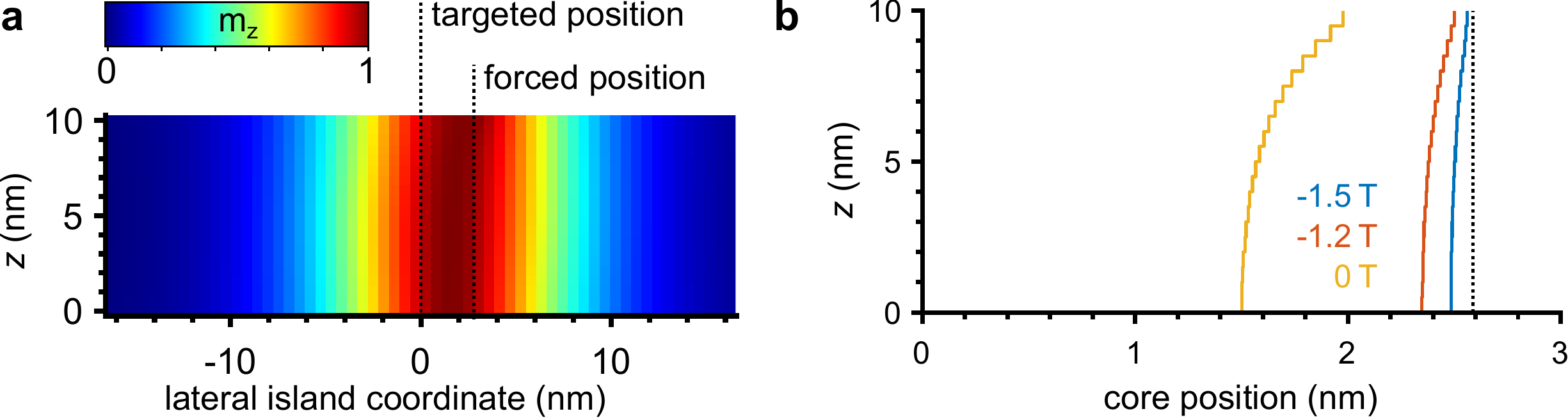}
\caption{\label{figS5}
\textbf{Core bending in Fe island.}
\textbf{a}, Cross-sectional view of perpendicular magnetization $m_z$ at $B_\perp=0$\,T recorded for a plane that is 1\,nm offset from the vortex core center. The circular island has thickness of 10\,nm and diameter of 280\,nm. The core is forced away from the island center by fixing $m_z$ in $4\times 4$ surface cells at $2.6$\,nm.
\textbf{b}, Core position ($m_z$ maximum) in each of the 20 layers of the simulated island for $B_\perp$ as labeled. The dotted black line depicts the position of frozen $m_z$.
%The core position is evaluated offset to the diagonal plane $y=x$, since the forced displacement by freezing $4\times4$ surface cells in the surface layer deforms the core significantly in that region.
}
\end{figure}

We analyze micromagnetic simulations with vortex cores shifted from the island center. The shift is achieved by fixing  $m_z$ within $4\times4$ surface cells offset from the island center. The fixed $m_z$ values are set to the values that are found in the core center for simulations without defects. The resulting cross section of $m_z$ through the island (Fig. \ref{figS5}a) is analyzed. We use cross sections slightly offset from the island center to avoid the cells with artificially fixed $m_z$.
Figure \ref{figS5}b shows deduced core positions ($m_z$ maxima) evaluated for each layer separately.
The core at $B_\perp = 0$\,T (-1.2\,T, -1.5\,T) is bent by 30\,\% (6.3\,\%, 2.7\,\%) of the average displacement from the island center. The bending at $B_\perp \neq 0$\,T, where we observe pinning in the experiment, is well below 10\% and, hence, barely changes the pinning energy. Such core bending is anyway included in our micromagnetic simulations of $u_{\rm{exch}}$ (Fig. 3g of the main text) and in the calculation of the parabolic potential $E_{\rm{flex}}$ for moving the vortex core away from its target (Fig.\,\ref{figS3}a, Fig.\,\ref{figS8}a).

\clearpage

%%%%%%%%%%%%%%%%%%%%%%%%%%%%%%%%%%%%%%%%%%%%%%%%%%%%%%%%%%%%%%%%%%%%%%%%%%%%%%%%%%%%

\subsection*{S6: Core position error using $dI/dV$ data at fixed position and varying $B_\parallel$}
\label{sec:immutable}
\begin{figure}
\includegraphics[width=16cm]{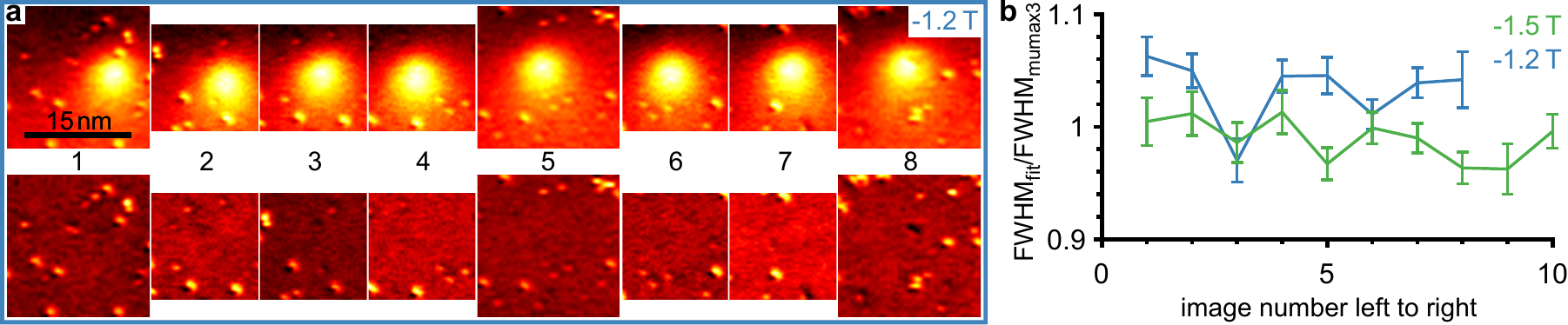}
\caption{\label{figS6}
\textbf{Core diameters for various core positions.}
\textbf{a}, Top row: d$I$/d$V$ images at $B_\perp = -1.2$\,T used to determine the 8 core positions marked in Fig. 2b of the main text.
The images are fitted as described in section S3, resulting in the residuals as shown in the bottom row.
The contrast of each image is scaled differently to optimize visibility.
\textbf{b}, FWHM of the $m_z$ distribution deduced via fit of the images of a (blue) and of 10 images recorded at $B_\perp = -1.5$\,T with core positions as marked in  Fig. 2b of the main text (green). The FWHM obtained from the fit is scaled to the FWHM of a simulation without defects (called stretch factor in Fig.\,\ref{figS7}) to ease comparison of the data at different $B_\perp$. The error bars correspond to 95\,\% confidence interval.
}
\end{figure}

In Fig.\,2 of the main text, we deduced the vortex core position from measuring $dI/dV$ at fixed tip position, while varying ${\bm B}_\parallel$. This assumes a rigid vortex motion along a straight path. The assumption implies errors, since the vortex core shape could change by interactions with defects and the core is displaced from the straight path due to defect pinning as visible in Fig.\,2b of the main text. These errors are discussed in the following.

To quantify the change of core shape, we analyze the core images along the core path of Fig.\,2b of the main text (Fig. \ref{figS5}a).
The FWHM of $m_z$ distribution is deduced from core fitting as described in section S3 .
It is displayed in Fig. \ref{figS5}b varying by about $\pm5\,\%$ without any obvious trend within the error bars from the fitting procedure. Hence, core shape modifications during pinning are below $5\,\%$.
This value is regarded as error  for the link between measured d$I$/d$V(B_\parallel)$ and core displacement (Fig.\,2 of main text).

Moreover, the core path is deflected from the straight path by the defects. It exhibits RMS deviations perpendicular to the target path up to 1.3\,nm (Fig.\,2b of main text). This implies two systematic errors.
First, the path gets longer by the zigzag motion such that $\chi_{\rm{pinned}}$ is underestimated by assuming a straight path. This error is estimated straightforwardly by using the measured path of Fig.\,2b of the main text. The real path is
by 5\,\% (0.3\,\%) longer than the straight path at $B_\perp= -1.5$\,T (-1.2\,T).
% Previous statement was "up to 14 % (1 %)", which is not sensible , because \chi-Pinned is later only given as an averaged value, of which the error has to be determined.
For the estimate, we measure the largest angle between target path and direct lines between adjacent core positions to be $\sim$30\textdegree  ($\sim$7\textdegree) at $B_\perp= -1.5$\,T (-1.2\,T) (Fig.\,2b, main text) and assume a normal distribution of such angles between adjacent core positions up to the maximum angle.

Second, perpendicular motion changes the sensed core magnetization at fixed tip position since the tip probes another part of the vortex. This error largely disappears for multiple pinning sites, since it either enhances or decreases $\chi_{\rm{pinned}}$ by corresponding changes of $dI/dV$ depending on the individual core center position with respect to the tip and the target path.

\clearpage

%%%%%%%%%%%%%%%%%%%%%%%%%%%%%%%%%%%%%%%%%%%%%%%%%%%%%%%%%%%%%%%%%%%%%%%%%%%%%%%%%%%%

\subsection*{S7: Approximations for micromagnetic simulations}
\label{sec:mumax approx}

For simulated sweeps of ${\bm B}_\parallel$, as employed for Fig. 2f-h, Fig. 3e, and Fig. 4b of the main text, two approximations are used to reduce computational time. They are validated in the following.

As first approximation, instead of sweeping $\bm{B}_\parallel$, we shift the defect by $-\chi_{\rm{free}}(B_\perp)\cdot {\bm B}_\parallel$
through the vortex core.
This requires that $E_\mathrm{flex} ({\bm r}_\mathrm {vortex}-{\bm r}_\mathrm{target})$, the displacement energy of the vortex around a target position $\bm{r}_{\rm{target}} = (x_{\rm{t}},y_{\rm{t}})$, does not depend on $\bm{r}_{\rm{target}}$.

\begin{figure}
\includegraphics[width=14cm]{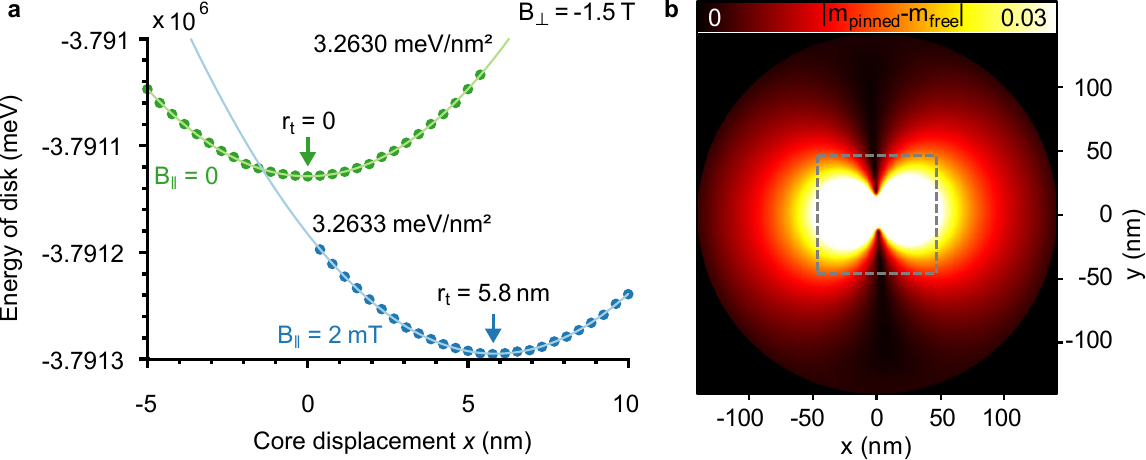}
\caption{\label{figS3}
\textbf{Validation of approximations in micromagnetic simulations.}
\textbf{a}, Micromagnetically calculated potential energy of vortex core displacement for $B_\perp = -1.5$\,T and $B_y=0$\,T (green dots)  as well as $B_y=2$\,mT (blue dots). The curvatures of the parabolic fits (solid lines) and the equilibrium positions $r_t$ are marked.
The micromagnetic simulations consider a cylindrical island of 280\,nm diameter and 10\,nm height discretized in cells of  $0.38\times0.38\times10$\,nm$^3$. Displacements are realized by fixing one cell to $m_z=1$ away from the target position ${\bm r}_{\rm{target}}$.
\textbf{b}, Absolute difference of $m_z$ between an unpinned vortex core and a core pinned at ${\bm r}_{\rm{vortex}}= (3.6, 0)$\,nm away from the island center ($B_\perp = -1.5$\,T).
Same island size and cell size as in a.
Only the area in the dashed box is used for full simulations of vortex-defect-interactions, while the remaining area is approximated by a demagnetization field independent of core position.
The normalized magnetization difference within the dashed grey box reaches up to 1.8, while it is below 3 \% outside of the box \cite{thesis_holl_2018}.
}
\end{figure}

To show this, we simulate $E_\mathrm{flex}$  for $\bm{r}_{\rm{target}}$ either located in the center of the island or offset from it (main text, methods).
We employ a grid with one cell in vertical direction for the sake of simplicity such that the core displacement is accomplished by a single cell of fixed $m_z=1$ located away from ${\bm r}_{\rm{target}}$.
It turned out that $E_\mathrm{flex}$ remained parabolic at all relevant distances of $\bm{r}_{\rm{target}}$ up to 30\,nm from the center of the island. Fig. \ref{figS3}a displays the micromagnetically calculated $E_{\rm{flex}}({\bm r}_\mathrm{vortex}-{\bm r}_\mathrm{target})$ for different $\bm{r}_{\rm{target}}$ along the target path in comparison with parabolic fits showcasing the nice agreement.
The curvature of the parabola changed by 0.01\,\% (10 \%) for distances of 5\,nm (30\,nm) from the island center.
We conclude that the displacement of the vortex mostly depends on the relative distance to the defect, but only marginally on the absolute position of the core within the island.
Hence, moving the defect instead of the vortex core is a reasonable approximation to deduce $\chi_\mathrm{pinned}$ (Fig. 2f-h and Fig. 3e of the main text). Note that Fig. 2f-h of the main text cover only $\pm 8$\,nm such that the curvature error is well below 1\%.

This agreement also justifies the assumption of a paraboloid for $E_{\rm{flex}}({\bm r}_\mathrm{vortex}-{\bm r}_\mathrm{target})$ for the simulation of core movement in the disorder potential as shown in Fig. 4b of the main text.
Deviations from the paraboloid in the direction perpendicular to the target path are even smaller, since the effective magnetization around the vortex is even less changed.

The independence of $E_{\rm{flex}}({\bm r}_\mathrm{vortex}-{\bm r}_\mathrm{target})$ from $\bm{r}_{\rm{target}}$ is corroborated by a simplified analytic model assuming a rigid movement of vortex magnetization by ${\bm B}_\parallel$ \cite{Rahm2004}. This employs the magnetic displacement model for a magnetic cylinder discussed in the main text with potential energy
$E(\bm{r}, \bm{B}_\parallel) = \frac{1}{2} k (x^2+y^2)-k\chi_{\rm{free}} (B_y x+B_x y)$. The equation can be rewritten as $E(\bm{r}) = \frac{1}{2} k ((x-x_{\rm{t}})^2+(y-y_{\rm{t}})^2) + \frac{1}{2} k (x_{\rm{t}}^2+y_{\rm{t}})^2$ with $x_{\rm{t}}=\chi_{\rm{free}}\cdot B_y$ and $y_{\rm{t}}=\chi_{\rm{free}}\cdot B_x$.
Hence, moving ${\bm r}_{\rm{target}}$ on a circular island leads only to an offset in potential energy (second term), but does not affect the potential curvature $k$ or the potential shape.

As second approximation, we crop the simulation area to $256\times256\times20$ cells and add the demagnetization field of the missing exterior by hand leading to an effective magnetic field ${\bm B}_{\rm{eff}} ({\bm r})={\bm B}_\perp+{\bm B}_{\rm{demag, exterior}}({\bm r})$.
${\bm B}_{\rm{demag, exterior}}({\bm r})$ is calculated once for an unperturbed vortex without defects at ${\bm B}_\parallel={\bm 0}$\,T and is fixed afterwards for all other simulations.
This is possible, since we always use ${\bm B}_\parallel={\bm 0}$\,T and, thus, ${\bm r}_{\rm{target}}={\bm 0}$\,nm via the first approximation.
The small core displacement resulting from pinning forces by defects changes the magnetization only within the cropped area significantly.
Figure \ref{figS3}b shows the spatially resolved absolute difference in magnetization between a vortex core located at ${\bm r}_{\rm{vortex}} = {\bm r}_{\rm{target}}={\bm 0}$\,nm and a core moved by pinning to ${\bm r}_{\rm{vortex}} = (3.6, 0)$\,nm at $B_\perp = -1.5$\,T. This displacement is larger than any displacement observed experimentally due to defects.
The scaled magnetization $m_z$ outside the fully simulated area (gray box) varies  by less than 3\,\% strongly decaying away from the square. As shown in section\,\ref{sec:energy density}, the general influence of demagnetization on the vortex core energy is small. Hence,
the resulting error of using an unmodified ${\bm B}_{\rm{demag, exterior}}({\bm r})$ is likely negligible.

\clearpage

%%%%%%%%%%%%%%%%%%%%%%%%%%%%%%%%%%%%%%%%%%%%%%%%%%%%%%%%%%%%%%%%%%%%%%%%%%%%%%%%%%%%

\subsection*{S8: Approximations for core path simulation}
\label{sec:path approx}

To emulate the vortex core path at varying $\bm{B}_\parallel$,
we determine its lateral position by potential energy minimization within a potential landscape given by defects as described in the main text. The potential energy firstly consists of the potential $E_{\rm{flex}}(\bm{r}_{\rm{core}}-\bm{r}_{\rm{target}})$ describing the energy cost to move the core  away from its  target position $\bm{r}_{\rm{target}} (\bm{B}_\parallel)$ in the absence of defects. Secondly, the pinning potentials centered at each adsorbate  $E_{\rm{i,pin}}(\bm{r}_{\rm{vortex}}-\bm{r}_{\rm{i,adsorbate}})$ contribute to the potential energy. For both potential parts, we use approximations
that enable easier computation.

\begin{figure}
\includegraphics[width=16.4cm]{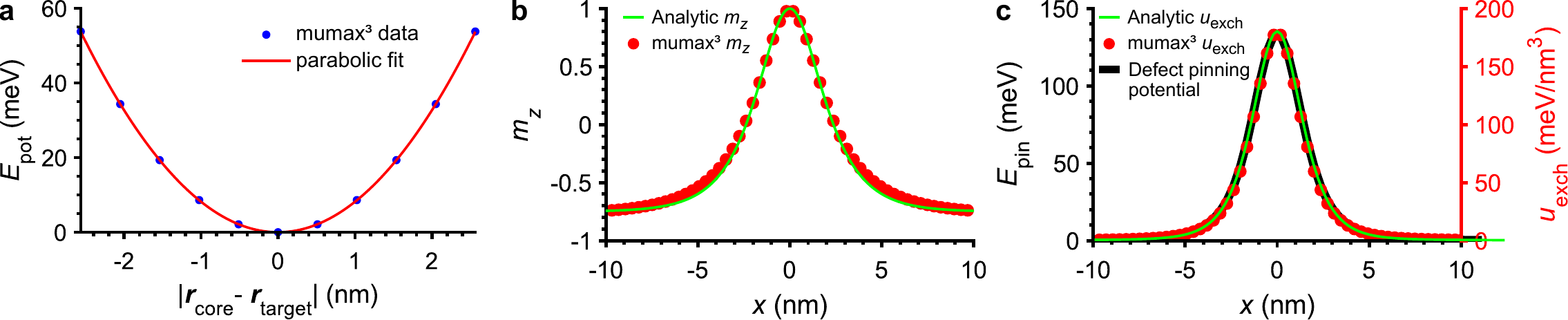}
\caption{\label{figS8}
\textbf{Approximations for core path simulation.}
\textbf{a}, Potential energy of vortex core without defects as a function of distance between core position $\bm{r}_{\rm{core}}$ and target position $\bm{r}_{\rm{target}}$, $B_\perp=-1.5$\,T. The simulation is based on a circular Fe islands (thickness: 10\,nm, diameter: 280\,nm) discretized into cells of size $0.364\times0.364\times0.5$\,nm$^3$.
For each data point, the core is forced away from $\bm{r}_{\rm{target}} = (0,0)$  by fixing $m_z$ in $4\times 4\times 1$ cells at the surface. The parabolic fit  (red line) yields negligible deviations from the data point of 0.03\,meV$_{\rm RMS}$.
\textbf{b}, $m_z$ profile of vortex core according to micromagnetic simulation by mumax$^3$ at $B_\perp=-1.5$\,T and to the analytic description of eq.\,(\ref{eq:mz}) with FWHM$=4.2$\,nm and $a=-0.8$.
\textbf{c}, Vortex core exchange energy density $u_{\rm{exch}}$  from the analytic description (eq.\,\ref{eq:uex}) with same parameters as in a and from mumax$^3$. The black line depicts the inverted pinning potential for a defect with quenched $A_{\rm{ex}}$ (Fig. 3g of main text), $B_\perp=-1.5$\,T.
}
\end{figure}

$E_{\rm{flex}}$ is deduced from forcing the vortex core away from $\bm{r}_{\rm{target}}$. Therefore, $m_z$ is fixed within $4 \times 4 \times 1$ simulation cells at the surface positioned away from $\bm{r}_{\rm{target}}$ to the $m_z$ values of a defect-free vortex core center. Subsequently, the vortex energy at the resulting core position is calculated. This mimics forcing the core away from $\bm{r}_{\rm{target}}$ by a defect. Such movement differs from movements via $\bm{B}_\parallel$ regarding the change of magnetization in the surrounding of the core. The required unphysical area of fixed magnetization
barely changes the vortex energy.
To estimate the corresponding error, we employed a second relaxation step while fixing the magnetization obtained from the first relaxation in all cells except of the priorily fixed ones and one additional ring of cells surrounding them.
For the largest core displacement observed at $B_\perp=-1.5$\,T, the potential energy changes by only 1.8\,\% due to this second relaxation step. Hence, the energy error of fixing $m_z$ in a few cells is well below 2\,\%.
Afterwards, the resulting  $E_{\rm{flex}}(\bm{r}_{\rm{core}}-\bm{r}_{\rm{target}})$ is fitted by a parabola (Fig. \ref{figS8}a). The fit exhibits a negligible RMS deviation of 0.03\,meV to the micromagnetic data for the largest displacements observed experimentally. Thus, we used a parabola for $E_{\rm{flex}}(\bm{r}_{\rm{core}}-\bm{r}_{\rm{target}})$ further on.

For the pinning potentials $E_{\rm{i,pin}}(\bm{r}_{\rm{vortex}}-\bm{r}_{\rm{i,adsorbate}})$, identical for each $i$, we superposed a repelling Gaussian and the scaled exchange energy density $u_{\rm{exch}} (\bm{r})$ of the core as described in the main text.
To increase computational speed, we employ an analytic representation of $u_{\rm{exch}}(\bm{r})$, based on an analytic approximation of $m_z(\bm{r})$:
\begin{equation}
\label{eq:mz}
m_z(\bm{r}) = a + (1-a)/\cosh(2 \cdot {\rm{arcosh(2)}}\cdot r/\rm{FWHM})
\end{equation}
with $a$ being the magnitude of $m_z$ in the surrounding of the vortex core and the width of the core $\rm{FWHM}$.
This leads to
\begin{align}
\label{eq:uex}
u_{\rm{exch}}(\bm{r}) &= A_{\rm{ex}}\cdot\Big(&&\nabla \bm{m}(\bm{r})\Big)^2\\ \nonumber
  &= A_{\rm{ex}}\cdot\Big( && \big( -b\cdot{\rm{tanh}}(r/c)\cdot{\rm{sech}}(r/c)/c\big)^2\\ \nonumber
  &&&+\big( 1-(b\cdot{\rm{sech}}(r/c)+a)^2\big)/r^2\\ \nonumber
  &&&+\big(b\cdot{\rm{tanh}}(r/c)\cdot{\rm{sech}}(r/c)
  \cdot(b\cdot{\rm{sech}}(r/c)+a)/c/\sqrt{1-(b\cdot{\rm{sech}}(r/c)+a)^2} \big)^2\Big)
\end{align}
with $b=1-a$ and $c ={\rm{FWHM}}/(2\cdot\rm{acosh}(2))$.

Figure \ref{figS8}c compares $u_{\rm{exch}}(\bm{r})$ from mumax$^3$ with the analytic description as best fit by adapting $a$ and FWHM. Excellent agreement is achieved with rms deviation of 0.6\,meV only.
The comparison of $m_z$ profiles is shown in Fig. \ref{figS8}b.
The reversed pinning potential for a  defect with suppressed $A_{\rm{ex}}$ within $1.1\times1.1\times0.5$\,nm$^3$ (Fig. 3g of main text) is added to Fig. \ref{figS8}c. Obviously, the relatively small defect simply tracks $u_{\rm{exch}}(\bm{r})$ such that the scaled analytic $u_{\rm{exch}}(\bm{r})$ can be used to mimic the attractive part of the defect potential for the core path simulation.
\clearpage

%%%%%%%%%%%%%%%%%%%%%%%%%%%%%%%%%%%%%%%%%%%%%%%%%%%%%%%%%%%%%%%%%%%%%%%%%%%%%%%%%%%%
\subsection*{S9: Errors in core path simulation and deduced pinning potential}
\label{sec:path errors}

The most severe error in core path simulation results from the remaining uncertainty in the adaption of the core shape at a defect. As shown in Fig.\,\ref{figS6}b, the FWHM of the $m_z$ profile fluctuates by $\pm 5$\%. This translates via eqs. (\ref{eq:mz}) and (\ref{eq:uex}) (section\,\ref{sec:path approx}) to an error of $\pm 5$\% in the FWHM of $u_{\rm{exch}}$, hence, influencing the pinning potential analogously by construction. The other energy errors are significantly smaller, namely, the error due to determination of $E_{\rm{flex}}$ via fixing $m_z$ in $4\times 4 \times 1$ simulation cells ($\le 1.8$\%, section\,\ref{sec:path approx}), the error due to determination of $E_{\rm{flex}}$ by moving the defect instead of the vortex core ($<1$\,\%, section\,\ref{sec:mumax approx}), the error due to the parabolic fit of  $E_{\rm{flex}}$ ($< 0.1$\%, section\,\ref{sec:path approx}) and the error due to the cropping procedure (likely negligible, section\,\ref{sec:mumax approx}).

Another source of error is more difficult to quantify. It is given by uncertainties in the determined core positions that are non-linearly linked to the deduced defect potential. This includes the missing knowledge on the true target path due to the fact that start and end point of the path of the vortex core are influenced by defects, too. The adaption of these points in our fitting routine reveals deviations by $1-2$\,nm on the full length of 40\,nm in line with typical excursion lengths from the straight path due to defects.  A similar deviation results from the anticipated straight target path in an elliptic island being incorrect by $1-2$\,nm on the path of 40\,nm, too (section\,\ref{sec:elliptic}). Other position errors are much smaller such as uncertainties in core center positions deduced from the fitting of noisy images ($< 0.1$\,nm, section\,\ref{sec:corefit}), uncertainties in the overlap of adjacent images of the vortex core ($< 0.1$\,nm) and creep and drift effects within the images ($\sim 0.1$\,nm, \cite{mashoff2009}).

Importantly, the main errors can be improved, in principle, via reducing the defect density, such that the distance between defects is significantly larger than the core diameter. Then, the influence of a single defect on the core shape can be probed in detail and start and end points of the target path can be chosen far away from any defect. Subleading errors can be reduced by more elaborate micromagnetic simulations.

\clearpage

%%%%%%%%%%%%%%%%%%%%%%%%%%%%%%%%%%%%%%%%%%%%%%%%%%%%%%%%%%%%%%%%%%%%%%%%%%%%%%%%%%%%
\subsection*{S10: Ab-initio calculations}
\label{sec:DFT}

We performed ab-initio based calculations of Cr- and O-adatoms deposited on an Fe(110) surface using density functional theory (DFT) as implemented in the full-potential Korringa-Kohn-Rostoker Green function (KKR-GF) method \cite{Papanikolaou2002,thesis_bauer_2014}. Relativistic effects are taken into account via the scalar relativistic approach with the self-consistent inclusion of the spin orbit coupling as a perturbation.
The exchange correlation potential is treated in the local spin density approximation as parametrized by Vosko, Wilk and Nusair  \cite{Vosko1980}.
Instead of seeking for the wave function of the system, the KKR-GF method aims primarily at calculating the Green function using multiple scattering theory by solving the Dyson equation:
\begin{equation}
\mathcal{G} = \mathcal{G}_0 + \mathcal{G}_0\Delta V \mathcal{G}. \label{dyson}
\end{equation}
This enables, e.g., to describe impurities deposited on a pristine substrate using an embedding scheme. Indeed, the previous Dyson equation can be solved in real space by obtaining the Green function $\mathcal{G}$ of the investigated material by knowing the Green function $\mathcal{G}_0$ of the perfect Fe(110) substrate and $\Delta V$, the potential change induced by the adatom. Once the Green function is obtained, the electronic and magnetic properties are deduced by extracting, e.g., charge and spin densities, local density of states, and magnetic exchange interactions.

\begin{figure}
\includegraphics[width=16.4cm]{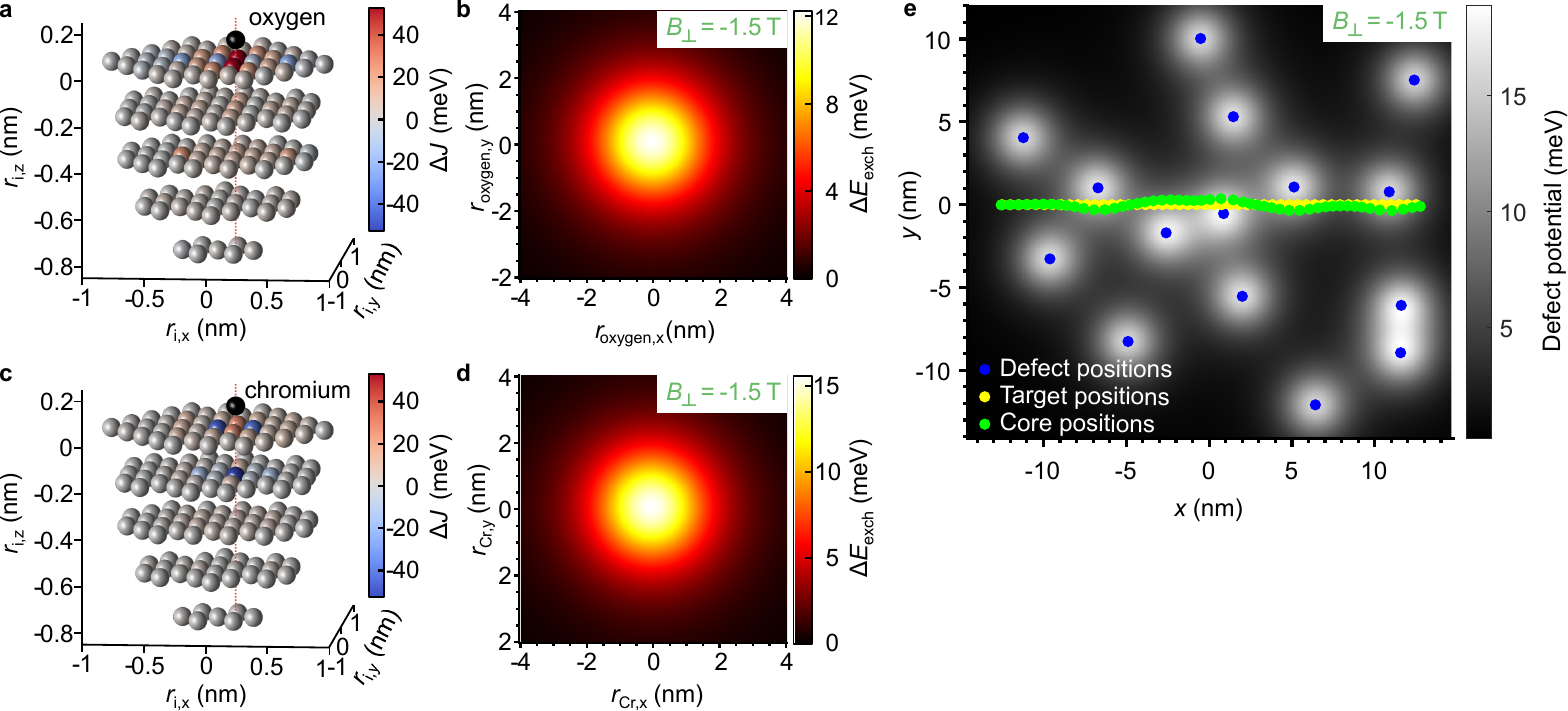}
\caption{\label{figS10}
\textbf{Ab-initio based vortex core energy around O and Cr adsorbates.}
\textbf{a}, Change of the site dependent $J_i=\sum_j J_{ij}/2$  for an Fe-cluster due to adding an O adsorbate (black) at a long bridge position. The difference $\Delta J_i=J_i^{\rm{with\hspace{0.5mm} O}}-J_i^{\rm{without\hspace{0.5mm} O}}$ is color coded on the grey spheres representing the Fe atoms, i.e., red (blue) color indicates a stronger (weaker) ferromagnetic coupling of the atom at $r_i$ to the other Fe atoms.
\textbf{b}, Resulting exchange energy potential of the vortex for varying vortex core position with respect to the O position, $B_\perp = -1.5$\,T. The potential is set to zero far away from the O atom. For each pixel of the potential, the magnetic moments \textbf{$\bm{m}_i (\bm{r}_i)$}
of a micromagnetically obtained vortex without defect are used to calculate $\Delta E_{\rm{exch}}=\sum_{i<j}(J_{ij}^{\rm{with\hspace{0.5mm} O}}-J_{ij}^{\rm{without\hspace{0.5mm} O}})(\bm{m}_i \cdot \bm{m}_j)$ for the respective vortex core center position with respect to the O position.
\textbf{c}, Analogous to a, but with Cr adsorbate.
\textbf{d}, Analogous to b, but with Cr adsorbate. Exchange coupling between Cr and the substrate atoms is taken into account.
\textbf{e}, Simulated vortex core path (green) at $B_\perp = -1.5$\,T employing 15 Cr defects that are randomly placed within $10\times10$\,nm$^2$ according to the defect density of the experiment.
The resulting disorder potential is displayed  as grey scale as deduced from superposing the defect potential of d for each adsorbate. The target path (yellow) consists of 50 equidistant positions along $y=0$.}
\end{figure}

The Fe(110) substrate with a lattice constant of $a_{\rm{lat}}=384$\,pm is simulated considering a slab containing 12 layers of Fe with enough vacuum layers surrounding it, six on each side of the slab. After relaxing the atomic positions at the surface, leading to values in accordance with \cite{Ossowski2015}, we solve the previous Dyson equation for a real-space impurity cluster. This cluster has a diameter of 6 lattice constants and consists of the adsorbate and 150 Fe atoms from the substrate (Fig. \ref{figS10}a/c). The adsorbates, O or Cr, are located in the long bridge position at a distance of 103\,pm above the surface as known for O \cite{Miyano1986,Getzlaff1999,Eder2001,Ossowski2015} and assumed to be identical for Cr.

Without the adsorbate, the average magnetic moment of the Fe atoms is 2.65 $\mu_\mathrm{B}$. With O (Cr), the closest Fe  moment decreases to 1.68\,$\mu_{\mathrm{B}}$ (0.68\,$\mu_{\mathrm{B}}$) while the substrate without considering the adsorbate experiences a cumulative reduction of the magnetization by 1.7\,$\mu_{\rm{B}}$ (4.6\,$\mu_{\mathrm{B}}$).

The change of the anisotropy due to the oxygen adsorbate was calculated by the energy difference $\Delta E_{\alpha-\beta} = (E_{\alpha}^{\rm{with\hspace{0.5mm} O}}-E_{\beta}^{\rm{with\hspace{0.5mm} O}})-(E_{\alpha}^{\rm{without\hspace{0.5mm} O}}-E_{\beta}^{\rm{without\hspace{0.5mm} O}})$, where $\alpha$ and $\beta$ denote the orientation of a ferromagnetic spin configuration along $[001]$ ($x$-axis), $[1\bar{1}0]$ ($y$-axis) or $[110]$ ($z$-axis). Hence, $\Delta E_{\alpha-\beta}>0$ indicates that the easy axis turns towards the $\beta$-direction when putting the O-adatom on top. The calculated values are $\Delta E_{x-z} = -0.68$\,meV, $\Delta E_{y-z} = -0.82$\,meV and $\Delta E_{x-y} = 0.13$\,meV, i.e., an in-plane magnetization along $y$ is favoured by the O-adatom. Importantly, these energies are much lower than the experimentally observed pinning energies ($\sim 100$\,meV) discarding any influence of the anisotropy energy on the pinning.

Utilizing a mapping procedure based on infinitesimal rotation of the magnetic moments \cite{Liechtenstein1984, Liechtenstein1987}, the magnetic exchange interactions $J_{ij}$ of an isotropic Heisenberg Hamiltonian $\mathcal{H}_\mathrm{exc} = -\sum_{i<j}J_{ij}\,  \bm{m}_i \cdot \bm{m}_j$, are extracted from the ab-initio calculations, where $\bm{m}_{i}$ and $\bm{m}_{j}$ are the unit vectors of the magnetic moments of the $i^{\rm{th}}$ and $j^{\rm{th}}$ atom, respectively.

A comparison of exchange parameters with and without adsorbate reveals that the cumulative exchange interaction is enhanced around both types of adatoms. Thus, we observe a global exchange stiffening.
Figure \ref{figS10}a shows a 3D map of the difference of the site dependent exchange parameter $\Delta J_i = J_i^{\mathrm{with\hspace{0.5mm} O}} - J_i^{\mathrm{, without\hspace{0.5mm} O}}$ where $J_i = \sum_j J_{ij}/2$. The Fe atoms nearest to the adsorbate along [1-10] ($y$-axis) exhibit a stiffening of the exchange interaction, while the exchange interaction along [001] ($x$-axis) gets weaker, but by a smaller amount.
The same is shown in Fig. \ref{figS10}c for the Cr adsorbate, where stiffening along [1-10] is weaker and weakening along [001] is more pronounced than for the O adsorbate.
The accumulated change in exchange energy amounts to $\Delta E_\mathrm{exch}=$\,217\,meV (86\,meV) for O (Cr) including the contribution of the Cr adatom of $-34$\,meV.
Since the exchange energy is increased in total (stiffening), a non-collinear magnetic texture as in the vortex core gains energy, if located away from the adsorbate, eventually leading to vortex core repulsion.

To reveal the interaction profile between adsorbates and vortex core, we employ $J_{ij}$ as obtained from DFT and calculate the exchange energy via $\mathcal{H}_\mathrm{exc}$ with the directions of the magnetic moments $\bm{m}_i$ set by the micromagnetically simulated vortex core profile. Changing the vortex core position with respect to the adsorbate reveals the interaction potentials as shown for $B_\perp = -1.5$\,T in Fig. \ref{figS10}b and d. The shape of the two potentials is identical
with slightly different amplitude of 12\,meV (15.5\,meV) for the O (Cr) adatom.
This amplitude is still an order of magnitude lower than in the experiment (Fig. 4c, main text).

Nevertheless, assuming the Cr induced interaction potential (Fig. \ref{figS10}d),
we simulated a vortex core path for randomly distributed Cr defects with density as in the experiment (Fig \ref{figS10}e). The simulation procedure is identical to the one employed for Fig. 4b of the main text. The resulting core path (green) at $B_\perp = -1.5$\,T deviates by up to 600\,pm from the straight target path (yellow). Such a deviation can be recorded by spin polarized STM and showcases that single adsorbates can alter the vortex path for a core size consisting of $\sim 10^4$ Fe atoms.

\begin{figure}
\includegraphics[width=16.4cm]{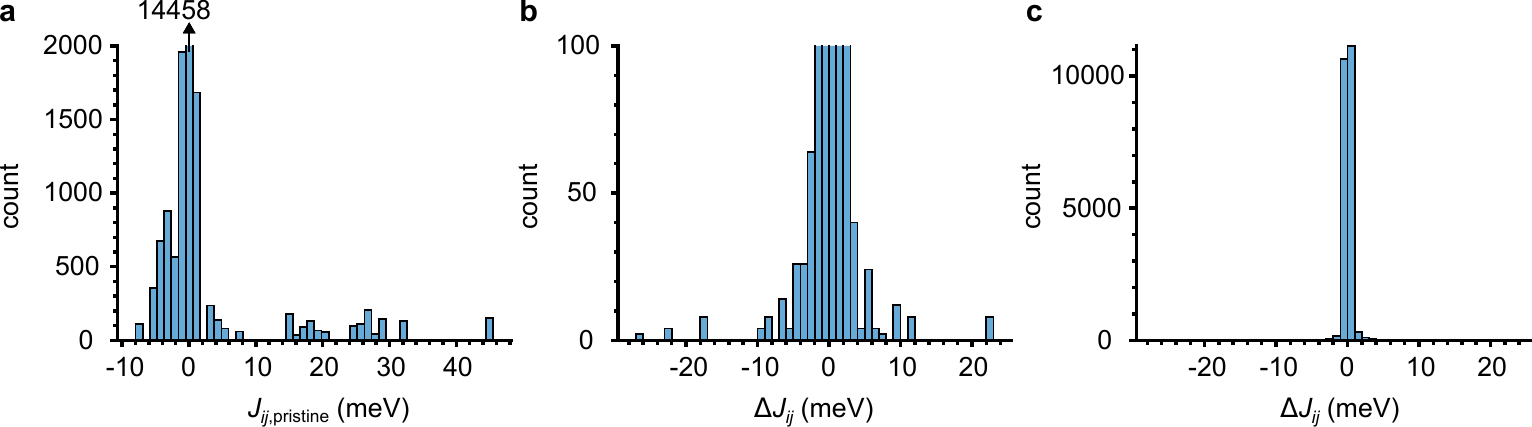}
\caption{\label{figS11}
\textbf{Histograms of exchange interactions between Fe atoms.}
\textbf{a}, Histogram of all $J_{ij}$ between the Fe atoms of the pristine substrate up to a distance of 6 lattice parameters around the site where O will be embedded.
\textbf{b}, Histogram of $\Delta J_{ij} = J_{ij}^\mathrm{with\hspace{0.5mm} O} - J_{ij}^\mathrm{without\hspace{0.5mm} O}$ , i.e., the  changes of $J_{ij}$ due to the O adsorbate for the same atoms as in a.
\textbf{c}, Same as b but displayed at a different scale.
}
\end{figure}

However, the much stronger excursions from the target path observed in the experiment can not be explained by this simulation.
One origin of the discrepancy could be different values of $J_{ij}$ than calculated via DFT.
Figure \ref{figS11} shows histograms of the exchange pa\-ra\-meter $J_{ij}$ for the Fe atoms of the pristine substrate  (Fig. \ref{figS11}a)  as well as of the change of the exchange parameters $\Delta J_{ij}$ due to adding an O adsorbate (Fig.\,\ref{figS11}b, c). The changes of $J_{ij}$ are partly as large as $J_{ij}$ itself. They, moreover, exhibit nearly as much reduction as increase of $J_{ij}$. In line, the accumulated $\sum \Delta J_{ij}=  217$\,meV amounts to only 10\% of the accumulated absolute energy change $\sum |\Delta J_{ij}|=2.5$\,eV.
This showcases that details in the interaction strengths $J_{ij}$ including sign changes can modify the accumulated exchange energy significantly via subtraction of two similarly large numbers.

Other possible origins of the discrepancy are already mentioned in the main text. Firstly, the structural position of the adsorbate might not be correctly described in the DFT calculations again changing $\Delta J_{ij}$ in detail. Secondly, the adsorbate might pinpoint to a particular strain field that might originate from the growth procedure and offers preferential adsorption sites.

\clearpage

%%%%%%%%%%%%%%%%%%%%%%%%%%%%%%%%%%%%%%%%%%%%%%%%%%%%%%%%%%%%%%%%%%%%%%%%%%%%%%%%%%%%

\subsection*{S11: Supplemental videos}
\label{sec:videos}

Supplemental video 1 consists of 45 d$I$/d$V$-images recorded at $B_\perp=-1.5$\,T, while moving the vortex core by 44 equidistant $\bm{B}_\parallel$ steps with $\Delta \bm{B}_\parallel=(136,-227)$\,$\upmu$T. These images are also used to determine the core positions shown in Fig. 1f of the main text.
Each $dI/dV$ image covers an area of $15\times15$\,nm$^2$.  Experimentally, $60\times60$ pixels are recorded at $V = -2$\,V, $I = 1$\,nA and modulation voltage of 50\,mV$_{\rm{RMS}}$. To optimize visibility, additional interpolated pixels are displayed in the movies. The scan frame center is moved linearly between adjacent images by a vector deduced from centering the core in initial and final image.
Supplemental video 2 shows the same data in different color scale and overlaid on a separately measured topography of the whole area. Here, the d$I$/d$V$-images are displayed after subtracting the contrast originating from in-plane magnetization and multiplying the image with a Gaussian intensity profile as described in section\,\ref{sec:corefit}.
Additional minor shear and stretch transformations by $\sim 1$\% are applied to remove the effects of piezo creep.

%\textbf{Data availability.}
%The data that supports the plots within this paper and other findings of this study are available from the corresponding author upon request.

%%%%%%%%%%%%%%%%%%%%%%%%%%%%%%%%%%%%%%%%%%%%%%%%
% SUPPLEMENT
%%%%%%%%%%%%%%%%%%%%%%%%%%%%%%%%%%%%%%%%%%%%%%%%

\end{document}